\documentclass[aps,pra,twocolumn,superscriptaddress]{revtex4-2}

\usepackage[english]{babel}
\usepackage{amsmath}
\usepackage{amsthm}
\usepackage{amssymb}
\usepackage{mathrsfs}
\usepackage{booktabs}
\usepackage{color}
\usepackage{hyperref}
\usepackage[ruled,vlined]{algorithm2e}
\usepackage{cases}
\usepackage{dblfloatfix}
\usepackage{multirow}
\usepackage{bbm}
\usepackage{enumitem}
\hypersetup{
  colorlinks   = true,  
  urlcolor     = blue,  
  linkcolor    = blue,  
  citecolor    = red    
}
\usepackage{mathtools}
\usepackage{natbib}
\usepackage{pgfplots}
\usepackage{subfigure}
\usepackage{url}

\newcommand{\di}{{\rm d}}

\newcommand{\F}{{\sf F}}

\newcommand{\V}{{\sf V}}
\newcommand{\Y}{{\sf Y}}

\newcommand{\z}{\mathbf{z}}

\newcommand{\bta}{\boldsymbol{\eta}}
\newcommand{\sfF}{\mathsf{F}}

\mathtoolsset{showonlyrefs}

\DeclareMathOperator{\Tr}{Tr}

\theoremstyle{definition}

\theoremstyle{remark}

\theoremstyle{plain}
\newtheorem*{theorem*}{Theorem}

\usepackage{scalerel,stackengine}
\stackMath
\newcommand\reallywidehat[1]{%
\savestack{\tmpbox}{\stretchto{%
  \scaleto{%
    \scalerel*[\wi\di thof{\ensuremath{#1}}]{\kern-.6pt\bigwedge\kern-.6pt}%
    {\rule[-\textheight/2]{1ex}{\textheight}}
  }{\textheight}%
}{0.5ex}}%
\stackon[1pt]{#1}{\tmpbox}%
}
\usepackage{comment}
\usepackage{float}

\SetArgSty{textnormal}

\makeatletter

\newenvironment{widetext2}{%
  \par\ignorespaces
  \setbox\widetext@top\vbox{%
   \vskip15\p@
   \hb@xt@\hsize{%
    \leaders\hrule\hfil
    \vrule\@height6\p@
   }%
   \vskip6\p@
  }%
  \setbox\widetext@bot\hb@xt@\hsize{%
    \vrule\@depth6\p@
    \leaders\hrule\hfil
  }%
  \onecolumngrid
  \let\set@footnotewidth\set@footnotewidth@ii
}{%
  \par
  \twocolumngrid\global\@ignoretrue
  \@endpetrue
}%

\makeatother

\begin{document}

\title{Fidelity-Enhanced Variational Quantum Optimal Control}

\author{R.J.P.T. \surname{de Keijzer}}
\altaffiliation[Corresponding author: ]{r.j.p.t.d.keijzer@tue.nl }
\affiliation{Department of Applied Physics, Eindhoven University of Technology, P. O. Box 513, 5600 MB Eindhoven, The Netherlands}
\affiliation{Eindhoven Hendrik Casimir Institute, Eindhoven University of Technology, P. O. Box 513, 5600 MB Eindhoven, The Netherlands}

\author{L.Y. \surname{Visser}}
\author{O. \surname{Tse}}
\affiliation{Eindhoven Hendrik Casimir Institute, Eindhoven University of Technology, P. O. Box 513, 5600 MB Eindhoven, The Netherlands}
\affiliation{Department of Mathematics and Computer Science, Eindhoven University of Technology, P.~O.~Box 513, 5600 MB Eindhoven, The Netherlands}

\author{S.J.J.M.F. \surname{Kokkelmans}}
\affiliation{Department of Applied Physics, Eindhoven University of Technology, P. O. Box 513, 5600 MB Eindhoven, The Netherlands}
\affiliation{Eindhoven Hendrik Casimir Institute, Eindhoven University of Technology, P. O. Box 513, 5600 MB Eindhoven, The Netherlands}

\date{\today}

\begin{abstract}
Creating robust quantum operations is a major challenge in the current noisy intermediate-scale quantum computing era. Recently, the importance of noise-resilient control methods has become more pronounced in the field. Ordinarily, noisy quantum systems are described by the Lindblad equation. However, minimizing noise susceptibility using this equation has proven challenging because of its irreversibility. In this study, we propose a new method for creating robust pulses based on the \textit{stochastic Schrödinger equation}. This equation describes individual noise realizations under any colored noise process, contrary to the Lindblad equation, which describes mean system behavior under white noise. Using stochastic optimal control techniques, our method, Fidelity-Enhanced Variational Quantum Optimal Control (F-VQOC), is able to construct higher fidelity paths than its non-stochastic counterpart (VQOC). By accounting for both environmental noise sources as well as noise sources inherent to the control system, highly significant increases in fidelity are noted for both single and multiqubit state preparations.

\end{abstract}

\maketitle

\section{Introduction}
\label{sec:introduction}


In the noisy inter\-mediate-scale quantum (NISQ) era, quantum computers operate with a limited number of qubits that are highly susceptible to noise \cite{Preskill_2018}, stemming from sources like stimulated emission, measurement errors or control noise \cite{morgado,errors1,errors2}. To advance beyond this stage in terms of noise resilience and scalability, significant improvements in both hardware and control software are essential. One promising approach on the software level is to create quantum operations that are least susceptible to noise by refining the control function---or pulse---that determines the trajectory of qubit states within the Hilbert space. Pulse construction algorithms aim to discover an optimal control pulse that transitions a qubit from an initial state to a (possibly unknown) target state. This field of study has gained research attention in recent years \cite{statepreppulse,qocvqe1,qocvqe2} with a focus on error mitigation, that is, minimizing noise and errors affecting the final state. One of the simplest strategies is to construct time-optimal pulses that reduce the exposure of the system to noise by shortening the interaction time of the system with its environment, as demonstrated in \cite{jandura,timeoptimal2}.


More sophisticated approaches for noise mitigation have also emerged. For example, dynamic decoupling methods apply near-instantaneous pulses to counteract static system-environment interactions \cite{dd1,dd2}. Rather than directly countering the effects of noise, Refs.~\cite{madhav, kosut2022robust,robust2,robust3,Poggi2023UniversallyRQ} develop pulses that guide the qubit state along paths that are less susceptible to noise. These approaches introduce time-independent noise and use cost functions based on the fidelity with respect to a predefined gate. Of these approaches, Refs.~\cite{madhav,kosut2022robust} optimize gates using gradient estimation techniques, while Ref.~\cite{robust2} takes a supervised learning approach. Other research, such as Refs.~\cite{robust3,Poggi2023UniversallyRQ}, employ analytical methods like inversion and perturbation techniques. Additionally, recent work in Refs.~\cite{weidner2023robust,aroch2023mitigating} revolves around designing pulses to maximize fidelity while considering the Lindblad equation, using estimated gradients and a known target state. In this context, Ref.~\cite{PMPquantumcontrol} investigates the optimal control of the Lindblad equation via Pontryagin's maximum principle \cite{pontryagin} as an alternative.


\begin{figure}[b]
    \centering
\includegraphics[scale=0.41]{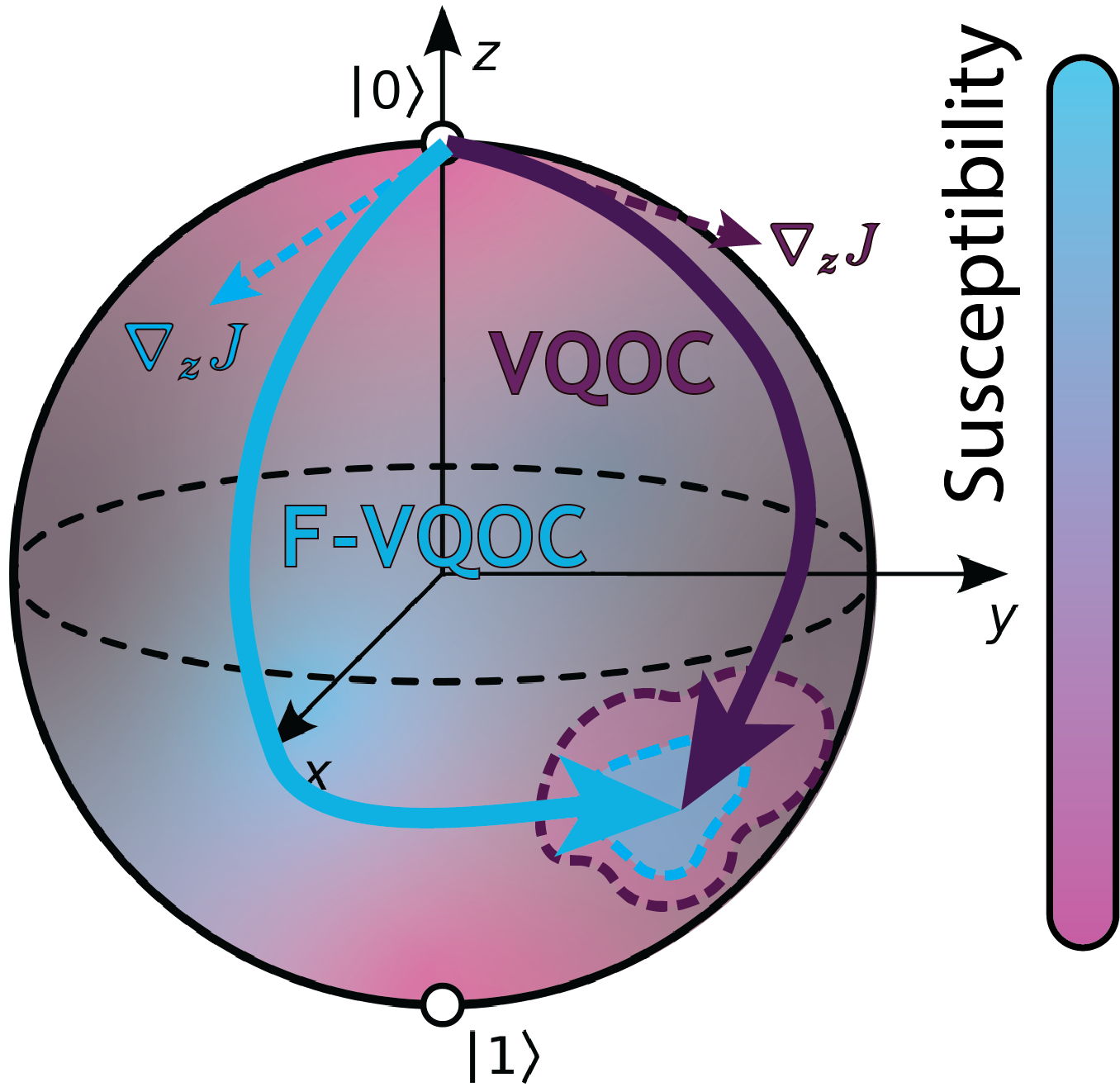}
    \caption{Illustration of a higher fidelity path (blue) on the Bloch sphere found by F-VQOC, compared to the low fidelity path (purple) found by standard VQOC. The incorporation of fidelity in the cost function leads to differing gradients $\nabla_zJ$, and subsequently lower error outcomes. Shade on Bloch sphere indicates areas of high/low susceptibility to noise.}
    \label{fig:abstract}
\end{figure}

Instead of just looking at the mean behavior of a quantum system described by the Lindblad equation, we consider a stochastic unraveling (or stochastic dilation) of this equation via the \textit{Stochastic Schrödinger Equation} (SSE) \cite{unravelling,Barchielli_2010,gardiner2004quantum}. The focus here is on noise sources driven by classical noise processes originating from the control pulses used for qubit state evolution, such as noise in laser intensity and frequency. Considering only classical noise simplifies the analysis and allows for a clearer understanding of how these common factors influence quantum systems. Yet, this method enables us to calculate the entire distribution of prepared states, moving beyond just the mean value \cite{ssepaper}. As an added complexity, we also consider noise intensity that scales with the amplitude of the control pulse, as would be the case in intensity noise in neutral atom systems \cite{parameters2} or flux noise in superconducting systems \cite{fluxnoise}. 

Since the SSE serves as a possible unraveling of the evolution of density matrices, (non-Markovian) Lindblad-type master equations can be derived from it, as shown in Refs.~\cite{ssepaper,semina,unraveltome}. In Ref.~\cite{wavefunctioncontrol}, optimal control techniques have been applied to create specific spatial wave functions through the SSE. Meanwhile, Ref.~\cite{mathematicalsse} rigorously shows the existence of solutions from a mathematical standpoint. To our knowledge, two other works, have utilized the SSE for enhancing fidelity in quantum operations.  Ref.~\cite{Ahn_2002} does this with a focus on quantum error correction rather than addressing noise mitigation, whereas Ref.~\cite{bkappen} uses path integral approaches.
 

In this work, we present a pulse construction method called \emph{Fidelity Enhanced} Variational Quantum Optimal Control (F-VQOC) based on the optimal control of the SSE, which we believe offers four key advantages: 
\begin{itemize}[itemsep=0pt,leftmargin=1em] 
    \item[--] The method is based on analytical gradients, allowing for the creation of high-fidelity pulses for systems with arbitrary numbers of qubits; 
    \item[--] It accommodates classical colored noise \cite{semimartingale}, enabling more realistic noise profiles on practical control systems based on power spectral densities \cite{psdwelch} compared to approaches assuming white noise; 
    \item[--] Fidelity optimization can be combined with unknown state preparations, such as ground state optimization in VQOC \cite{vqoc2} and other pulse-based Variational Quantum Algorithms (VQAs) \cite{qocvqe1,qocvqe2,GRAPE}, resulting in improved error rates; 
    \item[--] The algorithm generates pulses tailored to specific experimental setups by incorporating the characteristic noise operators of a system, allowing qubits to avoid noise-prone regions of the Hilbert space. 
\end{itemize}

The layout of this paper is as follows. Section~\ref{sec:VQOC} recalls (deterministic) VQOC, which is indifferent to noise. In Section~\ref{sec:SSE}, we detail the SSE and establish a framework that enables optimal control techniques. Section~\ref{sec:F-VQOC} provides the pulse optimization scheme for F-VQOC. In Section~\ref{sec:results}, we show initial results for our model on single and multiple qubit state and gate preparation, using classical white and colored noise. 

\section{Variational Quantum Optimal Control}
\label{sec:VQOC}

In variational quantum optimal control (VQOC) \cite{vqoc2}, a noiseless state $\phi$ \footnote{For readability purposes standard bra-ket notation is replaced by daggers to indicate conjugates, as in Refs.~\cite{semina,ssepaper}} evolves according to the Schr\"{o}dinger equation as
\begin{equation}
\label{eq:schrodinger}
    \frac{\di \phi_t}{\di t}=i H(t) \phi_t , \quad \phi(0)=\phi_0,
\end{equation}
where $\z:=(z^j)_{j\in J}$ is the vector of control pulses indexed by a finite set $J$ and 
\[
    H(t):=\sum_{j\in J} z_t^j H_j,\qquad t\ge 0,
\]
is the control Hamiltonian made up of a combination of Hamiltonians $H_j$, $j\in J$. The goal is to find optimal pulses by minimizing a cost function $J=J_1 + J_2$ with
\begin{equation}
\label{eq:vqoc}
    J_1(\z) := (\phi_T^\z)^\dagger H_{\text{targ}}\phi_T^\z, \qquad
    J_2(\z) := \frac{\lambda}{2}\|\z\|_2^2,
\end{equation}
where $\phi^\z_t$ is the unique solution corresponding to the Schr\"odinger equation \eqref{eq:schrodinger} for a specified control $\z$ \footnote{For notational purposes we will write $\phi_t$ for $\phi^\z_t$, where the dependence on $\z$ is implicit},  $H_{\text{targ}}$ is a given target Hamiltonian of which the ground state $\phi_{\text{targ}}$ is to be constructed, with ground state energy $E_{\text{targ}}=\phi_{\text{targ}}^\dagger H_{\text{targ}} \phi_{\text{targ}}$. The functional $J_1$ represents the energy of the state at time $t=T$, while the $J_2$ term penalizes for high-amplitude pulses.

\medskip
In VQOC \cite{vqoc2}, the Gâteaux derivative $\nabla_z J$ is calculated via adjoint methods and is determined as 
\begin{equation}
\label{eq:derivativeVQOC}
\begin{aligned}
    &\nabla J(\z)[\delta z^j] = -\lambda \int_{0}^T z^j_t\delta z^j_t\di t \\
    &-\int_{0}^T 2i\, (\phi_t^\z)^\dagger \Big[H_j^\dagger,\, \Gamma^\dagger(T,t) H_{\text{targ}} \Gamma(T,t)\Bigr]\phi_t^\z\delta z^j_t \di t,
\end{aligned}
\end{equation}
where $\Gamma(t,s):=U(t)U^\dagger(s)$, and $U$ is the unitary solution operator satisfying $\phi_t^\z=U(t)\phi_0$. Following the steepest descent along this gradient yields pulses that will map $\phi_0$ to a (local) minimizer of the cost function $J$ via \eqref{eq:schrodinger}.

For every ground state preparation problem, there are infinitely many possible paths $\phi_t$ connecting the initial state $\phi_0$ and the final state $\phi_T$. Figure.~\ref{fig:abstract} illustrates that some of these paths will be more susceptible to noise than others and thus will lead to worse final fidelity. If $\lambda>0$, the optimization algorithm will try to minimize the control function amplitude supplied to the system and thus reduce the influence of noise implicitly, as the signal-to-noise ratio of a control function is often constant over the signal strength \cite{stnratio}. However, during pulse construction, no direct attention is paid to the specific noise sources in the system. When implementing the constructed pulses in experimental systems, noise will continually supply errors into the system, which are especially relevant in the current NISQ era. This suggests a more sophisticated method, considering noise explicitly, with the goal of creating higher fidelity state preparations.

\section{Stochastic Schr\"{o}dinger Equation}
\label{sec:SSE}

To take into account the system's error sources during pulse construction, a descriptive model for the noise is necessary. These error sources might include spontaneous emission, dephasing \cite{dephasingnoise}, or, as analyzed exclusively in this paper, errors in the controls, e.g. caused by intensity or frequency noise in the lasers \cite{frequencynoise}. If one assumes a white noise profile for the control noise, the mean behavior of the system can be analyzed by the Lindblad equation, describing open quantum system evolution, and pulses that are more robust to these noises can be estimated numerically as in Refs.~\cite{weidner2023robust,aroch2023mitigating,PMPquantumcontrol}. However, we instead opt to employ the stochastic Schr\"{o}dinger equation (SSE) since it is more versatile in the noise profiles and provides information over the entire distribution of states, which can, in turn, be used for an analytical method to construct higher fidelity pulses. The SSE for the noisy state $\psi$ is given by
\begin{equation}
\label{eq:sse}
    \di \psi_t=iH\psi_t\, \di t-\frac{1}{2}\sum_{l} S_l^\dagger S_l \psi_t \di [X^l]_t+i\sum_l S_l\psi_t \di X^l_t, 
\end{equation}
where $S:=\{S_l\}_{l\in L}$ is a family of Hermitian noise operators indexed by a finite set $L$ and $X_l = (X^l_t)_{t\ge 0}$ are noise processes with finite quadratic variation $[X^l]_t = \gamma_l^2 t$, $\gamma_l>0$ \cite{semina}. This could for instance be white noise with $X^l_t=W^l_t$, but also Ornstein-Uhlenbeck (OU) noise \cite{ornstein} with
\begin{equation}
	\di X^l_t=-k_lX^l_t\,\di t+\gamma_l\,\di W^l_t,\quad k_l>0,
\end{equation}
or any noise sampled from an arbitrary power spectral density. As an illustrative case, we employ OU noise for this work, as it has a power spectral density that dominates in the lower frequencies, which is most common in real-world sources of control noise \cite{psdwelch}. In an ideal scenario, a pulse minimizing the ground state energy error 
\begin{equation}
\label{eq:error}
    J_\text{err}(\psi):=\mathbb{E}[\psi_T^\dagger H_{\text{targ}}\psi_T]-E_{\text{targ}}
\end{equation}
can be found. However, the quadratic dependence of this cost function on the stochastic wave function $\psi_T$ makes identifying an analytic expression for the gradient difficult. Instead, we adapt the VQOC problem \eqref{eq:vqoc} by adding a so-called \emph{fidelity regularizer} to the cost function, i.e.,
\begin{equation}
\label{eq:costfunction}
\begin{aligned}
    J &= J_1 + J_2 + J_3,\qquad\text{with} \\
    J_3(\z) &= -\mu\mathbb{E}\biggl[\sfF_T^\z+\nu\int_0^T \sfF_s^\z \,\di s \biggr],
\end{aligned}
\end{equation}
where $\mu,\nu>0$ and $\sfF_t^\z:=|(\phi_t^\z)^\dagger \psi_t^\z|^2$ is the fidelity of the stochastic state $\psi$ w.r.t.\ the deterministic state $\phi$. If $\nu>0$, the algorithm maximizes the fidelity over the entire path instead of just the end time. If $\mu=0$, the algorithm is inattentive to the fidelity and is identical to VQOC. In essence, F-VQOC is a fidelity-regularized version of VQOC. As a generale rule, we calculate $\nabla J_1(\z^{(0)})$ and $\nabla J_3(\z^{(0)})$ with $\z^{(0)}$ being the initial guess for the pulses, and choose $\mu=\|\nabla_{\z} J_1(\z^{(0)})\|/\|\nabla_{\z} J_3(\z^{(0)})\|$ to ensure both terms contribute equal orders of magnitude to the cost function.

In Sec.~\ref{sec:F-VQOC}, we provide the Gâteaux derivative of $J_3$, with the state evolving according to the stochastic differential equations of~\eqref{eq:sse}, for which we employ tools from stochastic optimal control \cite{Bensoussan_2018}. To this end, a system of evolution equations is constructed for the vector $\bta_t=(\phi_t^\dagger P_0 \psi_t, \phi_t^\dagger P_1 \psi_t,...,\phi_t^\dagger P_{4^N-1}\psi_t)^\top$ as performed in Ref.~\cite{ssepaper}, and described in detail in App.~\ref{app:matrices}. Here, $N$ is the number of qubits, and $P_i$ are the $4^N$ distinct Pauli matrices. Fixing $P_0=I$ ensures that the fidelity $\sfF_t$ can be equivalently expressed as
\[
    \sfF_t = \bta_t^\dagger \Lambda_0 \bta_t,\quad\text{with $\Lambda_0=[1,0,0,...,0][1,0,0,...,0]^\dagger$}.
\]
The system of equations for $\bta$ is given by
\begin{equation}
\label{eq:ssesystem}
    \di \bta_t 
    = g(\bta_t,\z(t))\, \di t + \sum_l f_l(\bta_t,\z(t))\, \di X_{l,t},
\end{equation}
with
\begin{align*}
    g(\bta,\z) &:= \sum_{j\in J} z^j A_j \bta -\frac{1}{2}\sum_{l\in L} \gamma_l^2 B_l^\dagger B_l \bta \\
    f_l(\bta,\z) &:= B_l \bta,\qquad l\in L,
\end{align*}
where the anti-Hermitian matrices $A_j$ and $B_j$ have elements 
\begin{equation}
\begin{aligned}
    (A_j)_{m,n}&=i\Tr[P_m[H_j,P_n]]=-\overline{(A_j)_{n,m}},\\
    (B_l)_{m,n}&=i\Tr[P_mP_nS_l]=-\overline{(B_l)_{n,m}}.
\end{aligned}    
\end{equation}
From \eqref{eq:ssesystem}, one can show that $\bta_t^\dagger \bta_t=\bta_0^\dagger \bta_0=2$ is a conserved quantity. 

\section{Fidelity Enhanced Variational Quantum Optimal Control}
\label{sec:F-VQOC}
Fidelity-enhanced Variational Quantum Optimal Control (F-VQOC) aims to minimize the cost function presented in \eqref{eq:costfunction} by employing an analytic gradient descent approach. To achieve this, we need to determine the Gâteaux derivative of the cost function. For the $J_1$ and $J_2$-terms outlined in~\eqref{eq:costfunction}, the Gâteaux derivative has been computed in Ref.~\cite{vqoc2}, as given in~\eqref{eq:derivativeVQOC}. For the fidelity term $J_3$, techniques derived from stochastic optimal control are required to compute the Gâteaux derivative. The calculations follow the general approach established in Ref.~\cite{Bensoussan_2018} and are detailed in App.~\ref{app:ssederivation}. The Gâteaux derivative of $J_3$ takes the form
\begin{equation}
    \nabla J_{3}(\z)[\delta z_j]=\mathbb{E}\left[ \int_0^T K_{z_j}(\bta_t, \z_t, p_t, r_t)\, \delta z_j(t)\, \di t\right],
\end{equation}
where
\begin{equation}
    K(\bta, z, p, r)=\Lambda_0 \bta_t+p^\dagger  g(\bta, z)+\sum_{l} r^\dagger_l f_l(\bta, z).
\end{equation}
Here, $p$ and $r$ are variables of backward stochastic differential equations, which are known to be difficult to solve \cite{backwardsde1}, both analytically and numerically \cite{backwardsde2}. We will examine two cases that are of physical relevance, where the Gâteaux derivative can be explicitly expressed using only forward stochastic differential equations, as elaborated in App.~\ref{app:ssederivation}. These being fixed noise, described in Sec.~\ref{sec:fixednoise}, and scaled noise as in Sec.~\ref{sec:scalednoise}. Lastly, in Sec.~\ref{sec:gate}, we will address the challenge of optimizing the fidelity over entire unitaries $I\rightarrow U_{\text{targ}}$ instead of a single state preparation $\phi_0\rightarrow \phi_{\text{targ}}$.

\subsection{Fixed noise}
\label{sec:fixednoise}
When the noise strength $\gamma>0$ is independent of the control functions---as is typical in scenarios involving external noise sources such as auxiliary electric fields \cite{electricfields,madhav}, spontaneous decays, or stimulated emission---the Gâteaux derivative can be expressed as
\begin{equation}
\begin{aligned}
        &\nabla J_{3}(\z)[\delta z_j]=\mathbb{E}\left[\int_0^T  \zeta_t \Psi_t A_j \bta_t\, \delta z_j(t)\,\di t\right],\\
        &\zeta_t :=\bta^\dagger_T \Lambda_0\Phi_T +\nu\int_t^T \bta_s^\dagger \Lambda_0 \Phi_s\, \di s, 
\end{aligned}
\end{equation}
where $\Phi$ is the solution operator of the stochastic system~\eqref{eq:ssesystem}, $\bta_t=\Phi_t x_0$, and $\Psi=\Phi^{-1}$. Both $\Phi$ and $\Psi$ follow forward stochastic differential equations and can thus be solved numerically using Monte Carlo simulation (see App.~\ref{app:stochint}), allowing for accurate and efficient approximation of the analytic expression for the gradient.

\subsection{Scaled noise}
\label{sec:scalednoise}
In realistic quantum systems, the noise associated with control pulses typically scales, most often linearly, with the amplitude of those pulses (e.g., \ amplitude and frequency noise). This scaling can be factored into the noise model~\eqref{eq:ssesystem} by considering noise processes $\di Y_{l,t}=\gamma_l \sqrt{|z_{c(l)}|}\, \di X_{l,t}$, which keeps the signal-to-noise ratio $\di [Y]_{l,t}/|z_{c(l)}|$ fixed. Here, $c$ maps the pulse $z_j(t)$ to which the noise profile $X_{l,t}$ scales, and $\gamma_l>0$ serves as the base noise level. The scaled noise SSE takes the form
\begin{equation}
\label{eq:ssesystemscaled}
\begin{aligned}
    \di \bta_t=&\sum_{j\in J} z_j(t) A_j \bta_t \di t-\frac{1}{2}\sum_l \gamma_{l}^2|z_{c(l)}(t)| B^\dagger_l B_l \bta_t \di [X]_{l} \\
    &+\sum_l \gamma_l\sqrt{|z_{c(l)}(t)|} B_l \bta_t \di X_{l,t}. 
\end{aligned}
\end{equation}
Assuming that $X_{l,t}$ is \emph{generated} \footnote{note that this does not imply $X_{l,t}=W_{l,t}$, instead $X_{l,t}$'s only random source is $W_{l,t}$, e.g. Ornstein-Uhlenbeck noise \cite{Barchielli_2010}} by a white noise process $W_{l,t}$, the derivative now takes the form
\begin{equation}
\begin{aligned}
        \nabla J_{3}(\z)[\delta z_j] &= \mathbb{E}\left[\int_0^T \zeta_t \Psi_t \bigg(A_j \bta_t\di t\right. \\
        &\hspace{-1em}\left.+\sum_{l|c(l)=j} \frac{1}{2} \frac{\gamma_l}{|z_j(t)|^{1/2}} B_l\bta_tdW_{l,t}\bigg)\delta z_j(t)\,\di t\right].
\end{aligned}
\end{equation}
For both fixed and scaled noise, the Gâteaux derivatives are determined by first generating noise realizations $X_{l,t}(\omega)$, and subsequently, calculating $x(\omega)$, $\Psi(\omega)$ and $\Phi(\omega)$ for each of these realizations. We then obtain the gradients by averaging these results. The pseudocode in Alg.~\ref{algo:QOCcontinuous2} shows an implementation of F-VQOC for fixed noise, which adapts easily to the scaled noise version.

\begin{algorithm}[h]
\SetAlgoLined
\SetInd{0.5em}{0.5em}
\SetKwData{Left}{left}\SetKwData{This}{this}\SetKwData{Up}{up}
\SetKwFunction{Union}{Union}\SetKwFunction{FindCompress}{FindCompress}
\SetKwInOut{Input}{input}\SetKwInOut{Output}{output}
\Input{$z^{(0)}$, $H_{\text{targ}}$, $\phi_0$, \#iter, \#trials}
\Output{$z^{(\text{\#iter})}, \mathbb{E}_\omega[\psi^\dagger_{\omega,T} H_{\text{targ}} \psi_{\omega,T}]$}
\BlankLine
\tcp{Pulse Optimization Procedure}
\For{$k=0$ \KwTo $ \text{\#iterations}$}{
calc $\nabla J_{1}, \nabla J_{2}$;\\
\For{$\omega=0$ \KwTo $ \text{\#trials}$}{
generate $X_{l,\omega}$;\\
calculate $\bta_\omega, \Phi_\omega, \Psi_\omega$;\\
$\nabla J^\omega_{3}=\zeta_\omega\Psi_\omega A_j \bta_\omega$;\\
}
$z^{(k+1)}_{j}=z^{(k)}_j-\alpha_k\bigl(\nabla J_{1}+\nabla J_{2}+\mathbb{E}[\nabla J^\omega_{3}]\bigr)$;
}
\tcp{Final Energy Determination}
\For{$\omega=0$ \KwTo $ \text{\#trials}$}{
generate $X_{\omega,l,t}$;\\
calc $U_\omega(z^{(\text{\#iter})})$;\\
}
return $z^{(\text{\#iter})}$, $\mathbb{E}_\omega[\phi^\dagger_0{U_\omega^\dagger}(T)H_{\text{targ}}U_\omega(T)\phi_0]$;\\
\caption{F-VQOC, fixed noise}\label{algo:QOCcontinuous2}
\end{algorithm}

\subsection{Gate Optimization}
\label{sec:gate}
Alongside state preparation (transitioning $\phi_0$ to $\phi_{\text{targ}}$), a key challenge is the construction of a complete unitary transformation $U_\text{targ}$. At first sight, one may not expect VQOC to benefit significantly from fidelity optimization because of the following naive intuition: rotating one state out of a noisy region would merely rotate another into that same noisy region. Yet, we will study the problem as it reveals surprising and insightful conclusions.

The evolution of both the noisy unitary $U$ and the noiseless unitary $V$ is characterized by
\begin{equation}
\label{eq:sseunitary}
\begin{aligned}
    &\di U_t=iH U_t\, \di t-\frac{1}{2}\sum_{l} S^\dagger_l S_l U_t \di [X]_{l,t}+i\sum_l S_l U_t \di X_{l,t},\\ 
    &\di V_t=iHV_t\, \di t,\quad U(0)=V(0)=I.
\end{aligned}
\end{equation}
For this problem, the ground state energy reads
\begin{equation}
    J_1(V)=-|\Tr[{U_\text{targ}^\dagger} V_T]|^2.
\end{equation}
The fidelity equivalent for gates $\sfF_t$ is given by an integral over all possible initial states $\phi_0$ distributed according to the Haar measure $\mu_{\text{Haar}}$ \cite{haarmeasure}:
\begin{equation}
\begin{aligned}
    J_3(U,V) &=\mathbb{E}\left[\int \phi_0^\dagger V_T^\dagger U_T\phi_0 \,\mu_{\text{Haar}}(\di \phi_0)\right]\\
    &= \mathbb{E}\left[    \left\langle 0\left|\int_{\bar{U}} \bar{U}^\dagger V_T^\dagger U_T \bar{U}\,\mu_{\text{Haar}}(\di \bar{U})\right| 0\right\rangle\right] \\ 
    &=\frac{1}{2^N}\mathbb{E}\Bigl[\Tr[V_T^\dagger U_T]\Bigr].
\end{aligned}
\end{equation}
where the last equality holds by the averaging property of the Haar measure, stating that, for any operator $O$,
\begin{equation}
    \int \bar{U}^\dagger O \bar{U} \mu_\text{Haar}(\di \bar{U})=\frac{1}{2^N}\Tr[O]I.
\end{equation}
The operator $Q_t:=V_t^\dagger U_t$ evolves according to 
\begin{equation}
\di Q_t=i[H,Q_t]\di t-\frac{1}{2}\sum_l\gamma^2S^\dagger_l S_lQ_t\di t + i\sum_lS_lQ_t \di X_{l,t},
\end{equation}
with initial data $Q(0)=I$. The continuous-time case ($\nu>0$) may be considered in a similar fashion.

The minimization for $J_3$, in this case, can be done using stochastic optimal control with 
\[
    \bta_t=\{\Tr[P_0Q_t],\Tr[P_1Q_t],...,\Tr[P_{4^N-1}Q_t]\}.
\]
Noticeably, this system evolves exactly as \eqref{eq:ssesystem}, but with $\bta_0=\{1,0,0,...,0\}$, and thus acts as if the initial states $\bta_0=\{1,\eta_{01},\ldots,\eta_{0N}\}$ in \eqref{eq:ssesystem}, where $\{\eta_{01},\ldots,\eta_{0N}\}$ are points on the $4^N-1$ dimensional sphere, are averaged out in the full gate optimization case.

Furthermore, considering white noise $X_{l,t}=\gamma_l W_{l,t}$ and Pauli-type noise operators $S_l^\dagger S_l=I$, we obtain
\begin{equation}
\frac{\di}{\di t}\mathbb{E}[\Tr[Q_t]]
    = -\frac{1}{2}\gamma^2\mathbb{E}[\Tr[S^\dagger S Q_t]]=-\frac{1}{2}\gamma^2\mathbb{E}[\Tr[Q_t]],
\end{equation}
indicating that the fidelity term $J_3(U,V)$ remains independent of the choice of pulses, as hypothesized at the beginning of this section. Nevertheless, when $S^\dagger S\neq I$ or in scenarios involving noise sources beyond white noise, one could expect to find a non-zero gradient of $J_3$.

\section{Results}\label{sec:results}
In this section, we present several examples to illustrate the performance of F-VQOC and provide a comparison with its deterministic counterpart, VQOC \cite{vqoc2}. In all cases, we will compare F-VQOC both with $\nu>0$ (for the continuous-time cost) and with $\nu=0$ (for the end-time cost) to VQOC ($\mu=0$, no SSE). Throughout this section, we incorporate relatively high noise strengths, characterized by signal-to-noise ratios on the order of $0.01$, which is high but not unreasonable for NISQ devices \cite{parameters1,parameters2,parameters3}. We do this to distinguish errors induced by the SSE from those arising from the pulse construction algorithm, which might struggle to find the exact minimizing pulse for the ground state problem. Furthermore, it showcases the effect F-VQOC can have on paths taken through Hilbert space. All averages are taken over $N=200$ runs. As a figure of merit, we will generally consider the relative increase in error $(J_\text{err}(\psi)_{\text{F-VQOC}}-J_\text{err}(\psi)_{\text{VQOC}})/J_\text{err}(\psi)_{\text{VQOC}}$ as in \eqref{eq:error}.

\subsection{Fixed noise}
In the case of fixed noise, we analyze a 1-qubit problem with the target Hamiltonian defined as $H_{\text{targ}}=-\sigma_Y$. Hence, the goal is to find control pulses that map the initial state, $	\phi_0=|0\rangle$, to the ground state, $	\phi_g=|+\rangle$. We assume full control over the Bloch sphere, i.e., we can apply $H_i=\sigma_i$ for $i\in\{X, Y, Z\}$. The control Hamiltonians are subjected to OU control noise, characterized by $S_i=H_i$, with values $(\gamma_{X},\gamma_{Y},\gamma_{Z})=(0.07,0.01,0.01)$ and $k_i=0.1$. To optimize the control, we implement the F-VQOC with the parameters $(\lambda, \mu, \nu)=(0.1, 250, 1)$ for the continuous cost function, while setting either $\nu=0$ or $\mu=0$ for the end cost and the VQOC, respectively. After 10 iterations, we set $\mu=0$ in all cases, aiming to hone in on achieving a high-fidelity path toward the approximate ground state. 

\begin{figure}[h]
    \centering
    \includegraphics[scale=0.65]{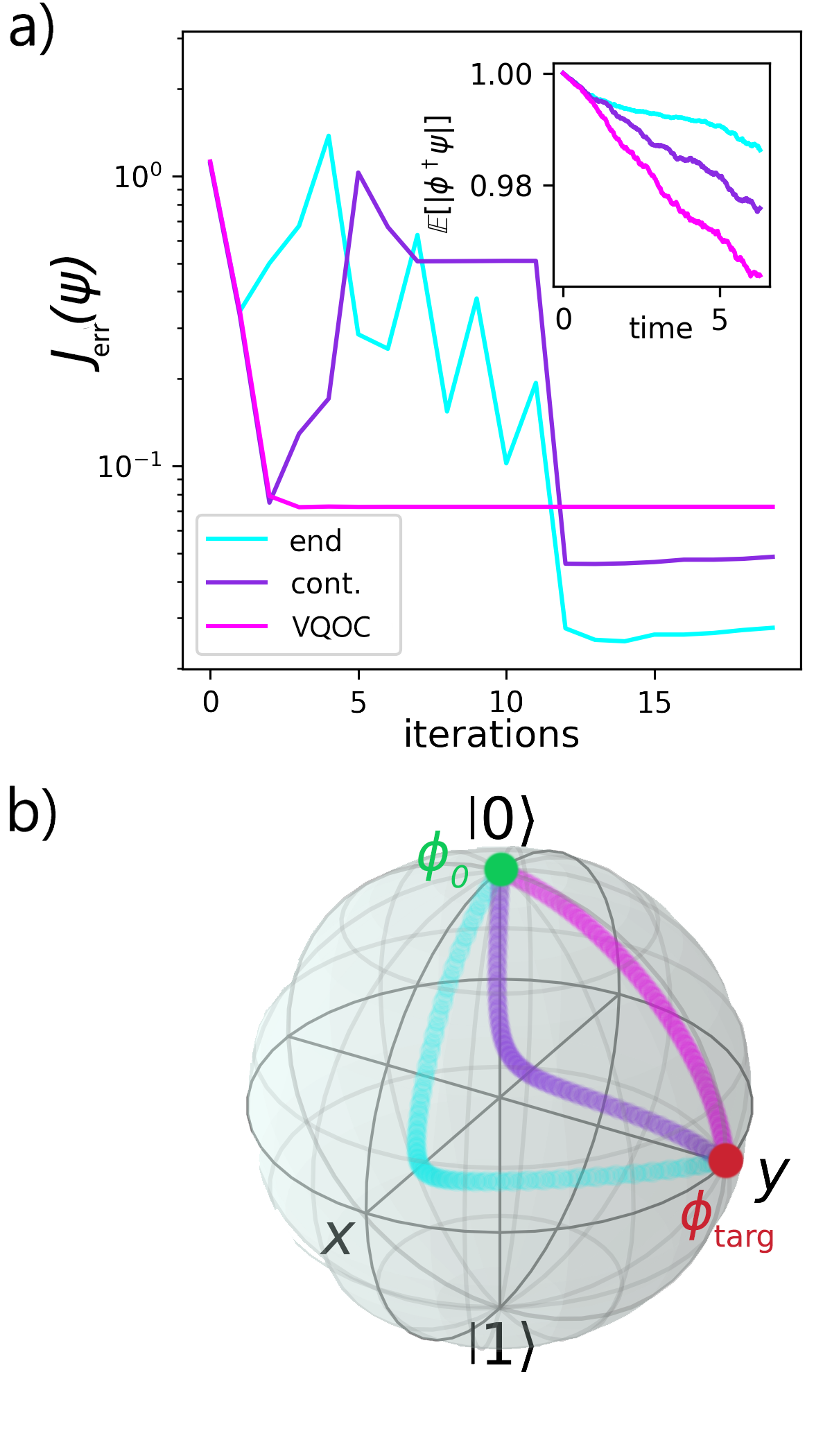}
    \caption{a) Empirical average energy found per iteration for F-VQOC vs. VQOC with $(\lambda,\mu,\nu)=(0.1,250,1)$ and $H_i=\sigma_i$, $S_i=H_i$ with OU noise with $(\gamma_{X},\gamma_{Y},\gamma_{Z})=(0.07,0.01,0.01)$ and $k_i=0.1$, for $i\in\{X,Y,Z\}$. b) Paths found on the Bloch sphere from $\phi_0=|0\rangle$ to $\phi_g=|+\rangle$ by the various methods.}
    \label{fig:onequbitfixedsingle}
\end{figure}

The energies and control paths identified in this problem are illustrated in Fig.~\ref{fig:onequbitfixedsingle}. One observes that the end-time cost demonstrates both higher fidelity and lower energy error compared to VQOC, with the continuous-time cost performing even better on both metrics. As anticipated, both methods converge towards the ground state after 10 iterations when $\mu=0$. The promising performance can be attributed to the paths found on the Bloch sphere, as shown in Fig.~\ref{fig:onequbitfixedsingle}b. Both F-VQOC methods successfully guide the qubit state towards the $\sigma_X$ eigenstates, where it is least susceptible to noise due to the condition $\gamma_X > \gamma_Y, \gamma_Z$.

\begin{figure}[b]
    \centering
    \includegraphics[width=0.9\linewidth]{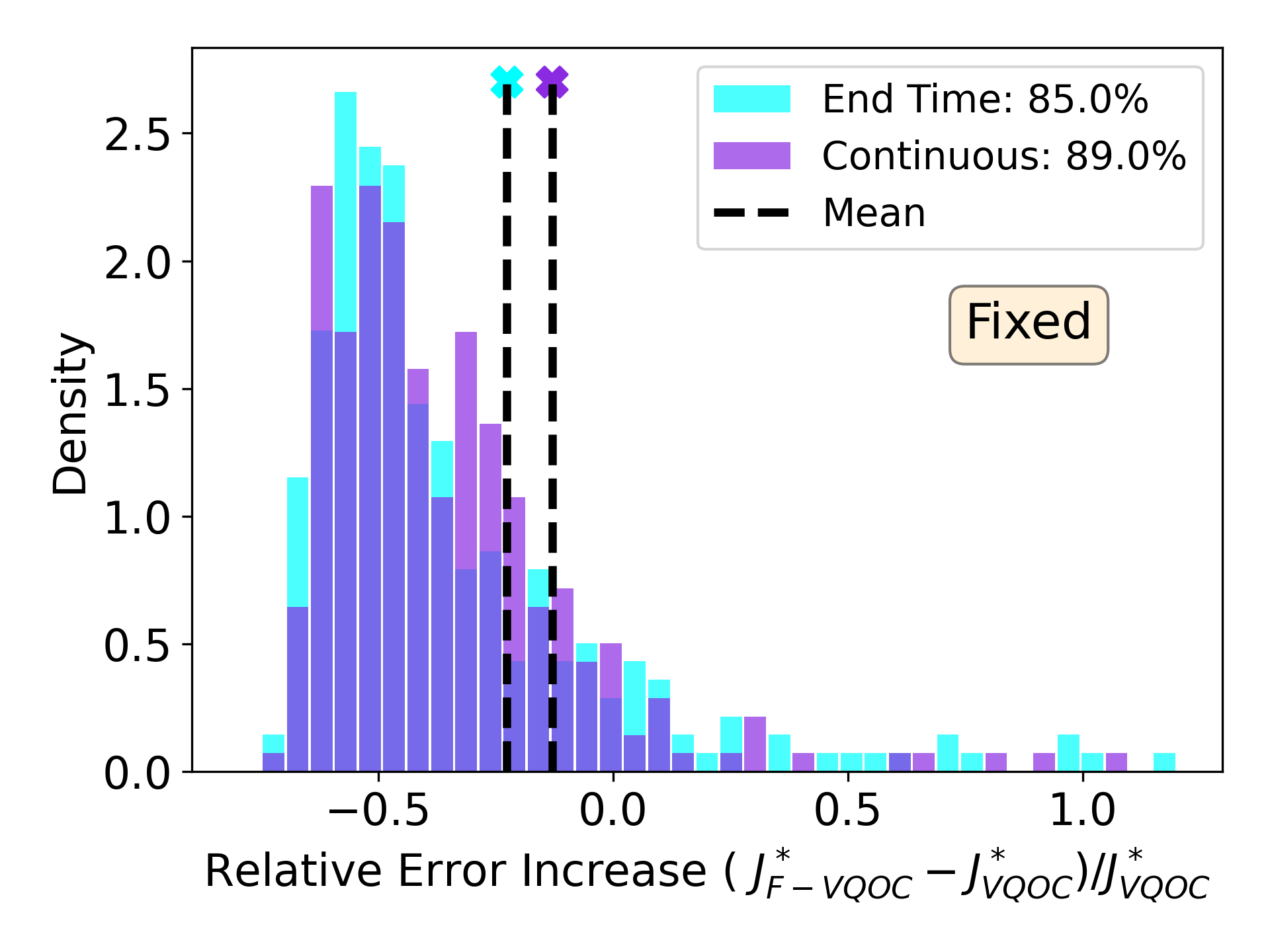}
    \caption{Distribution of relative ground state energy error increase F-VQOC vs.\ VQOC with $(\lambda,\mu,\nu)=(0.1,250,1)$ for random 1-qubit Hamiltonians $H_{\text{targ}}$, $H_i=S_i=\sigma_i$ with OU noise, $(\gamma_{X},\gamma_{Y},\gamma_{Z})\sim$ Unif$[0,0.1]^3$ and $k_i=0.1$, $i\in\{X,Y,Z\}$. Legend indicates the percentage of F-VQOC trials outperforming VQOC counterparts. Mean lines indicate mean relative error increase.}
    \label{fig:histogramfixed}
\end{figure}

In this specific example, we see that the choice of the target Hamiltonian and noise strengths leads to F-VQOC obtaining an intuitively higher fidelity path than VQOC. However, it is important to note that F-VQOC consistently outperforms VQOC, even in cases where the optimal control pulse is less apparent. To illustrate this point, we performed F-VQOC simulations using randomized 1-qubit target Hamiltonians paired with uniform random noise strengths $\gamma_i\in[0,0.1]$. Instead of a random target state and setting $H_{\text{targ}}=I-|\phi_{\text{targ}}\rangle\langle\phi_{\text{targ}}|$, we choose to randomize the entire Hamiltonian $H_{\text{targ}}$. This Hamiltonian has a broader spectrum, making the ground state harder to find \cite{spectrum}, showcasing the strength of our method. Figure~\ref{fig:histogramfixed} displays the distribution of relative error between F-VQOC and VQOC for the end and continuous-time costs for 300 of these random problem initializations. We observe similar performance improvements both for end and continuous-time costs, achieving an average reduction in error by 20\% and a decrease in error in 87\% of the cases examined. Notably, the relative error increases for F-VQOC seem to mostly happen when the absolute fidelity of VQOC is already high. 

It is worth mentioning that these results could likely be further improved by fine-tuning the regularization parameters within a fixed experimental setup, where the values of $\gamma_i$ are kept constant. Overall, these findings demonstrate that the F-VQOC method effectively balances the minimization of ground state energy while conserving fidelity, thereby enabling the identification of lower error paths through Hilbert space.

\subsection{Scaled noise}

For scaled noise, the experiment proceeds similarly, with modifications only to the regularization parameters set to $(\lambda,\mu,\nu)=(0.1,60,1)$, and $\mu=0$ is set only after 15 iterations, as this led to better convergence performance. Furthermore, the target Hamiltonian is specified as $H_{\text{targ}}=-|1\rangle\langle 1|$. 

\begin{figure}[h]
    \centering
    \includegraphics[scale=0.65]{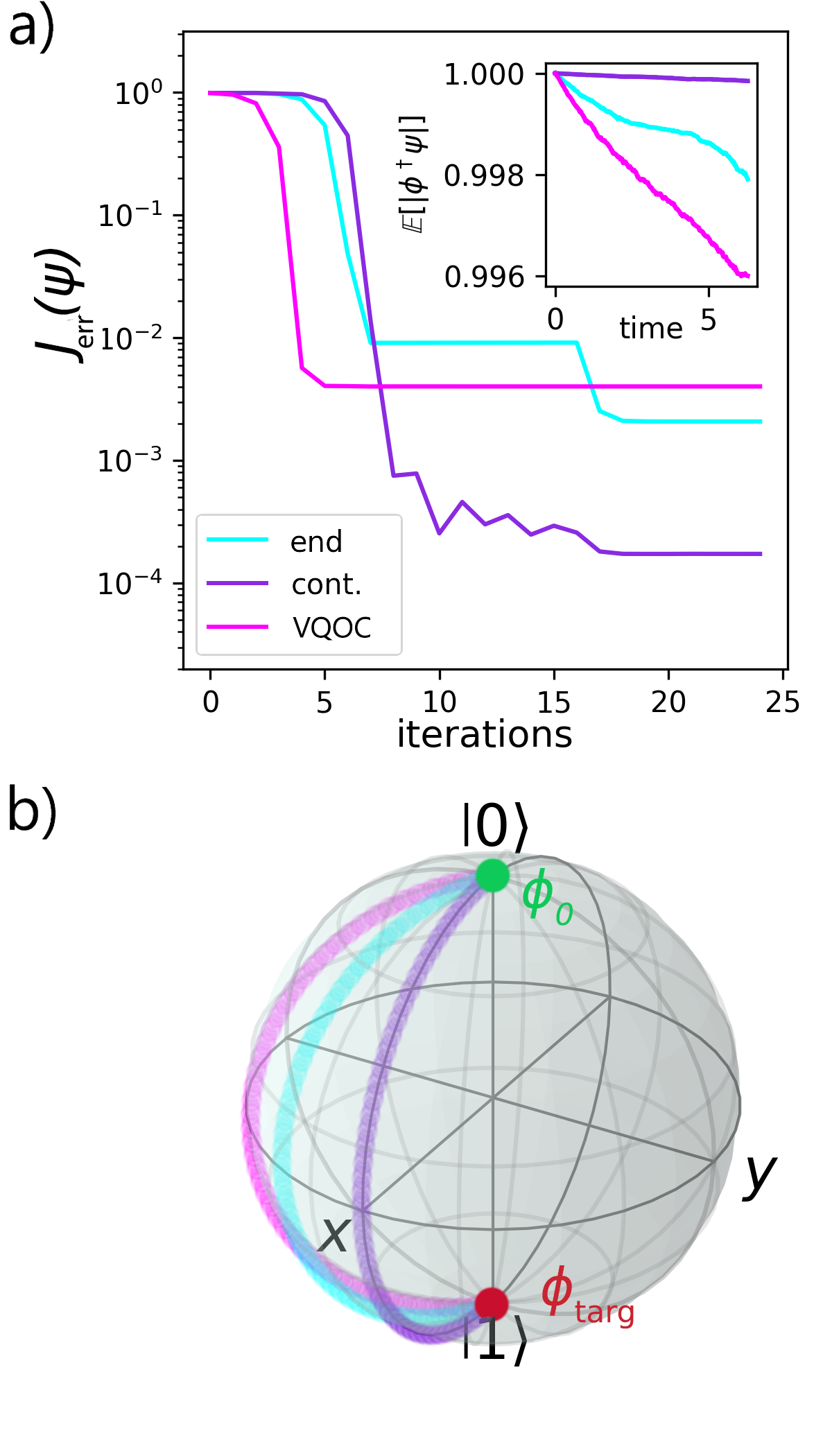}
    \caption{a) Average energy found per iteration for F-VQOC vs. VQOC with $(\lambda,\mu,\nu)=(0.1,60,1)$ and $H_i=\sigma_i$, $S_i=H_i$ with OU noise with $(\gamma_{X},\gamma_{Y},\gamma_{Z})=(0.07,0.01,0.01)$ and $k_i=0.1$, for $i\in\{X,Y,Z\}$. b) Paths found on the Bloch sphere from $\phi_0=|0\rangle$ to $\phi_g=|1\rangle$ by the various methods.}
    \label{fig:onequbitscaledsingle}
\end{figure}

Figure~\ref{fig:onequbitscaledsingle} shows the results of the algorithm addressing this problem. Here, VQOC drives $\sigma_X$ and $\sigma_Y$ simultaneously, as this leads to lower pulse norm $J_2=\|z\|_2$. Consistent with the findings of the fixed noise case, F-VQOC outperforms VQOC in both occasions by navigating a path closer to the noise immune eigenstates of $\sigma_X$. 

\begin{figure}[H]
    \centering
    \includegraphics[width=0.9\linewidth]{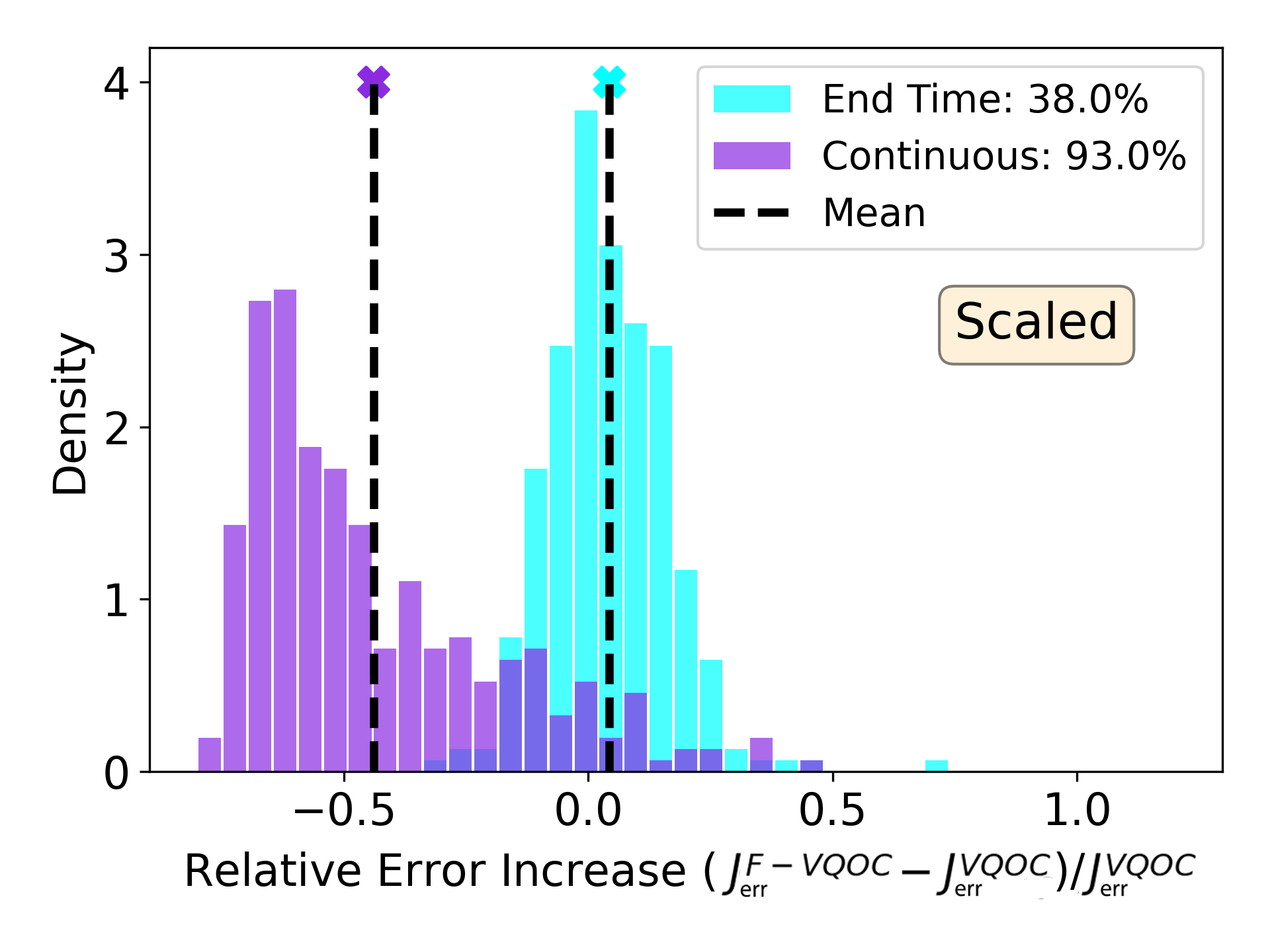}
    \caption{Relative ground state energy error increase density plots  F-VQOC vs.\ VQOC with $(\lambda,\mu,\nu)=(0.1,60,1)$ and $H_i=\sigma_i$, $S_i=H_i$ with OU noise with $(\gamma_{X},\gamma_{Y},\gamma_{Z})\sim$ Unif$[0,0.1]^3$ and $k_i=0.1$, for $i\in\{X,Y,Z\}$ and $H_{\text{targ}}$ a random single qubit Hermitian. Legend indicates what percentage of F-VQOC trials outperform their VQOC counterpart. Mean lines indicate mean relative error increase.}
    \label{fig:histogramscaled}
\end{figure}

Interestingly, as seen in Fig.~\ref{fig:histogramscaled}, while the end-time cost performs similarly to VQOC, the continuous-time cost significantly outperforms it in most of the randomized trials. We hypothesize that this behavior emerges from the scaling of noise with pulse strength, making the direction of the path critically important, rather than its strength. A continuous cost would then find a path that goes to a noise-insensitive area early in the evolution, contrasting with the delayed responses of an end-time cost. Another reason could be due to bad choices of the regularization parameters $(\lambda,\mu,\nu)$. However, variations of these parameters were tested, and initial findings suggest that this is not likely. 

\subsection{Gate Optimization}

We implement the gate construction method outlined in Sec.~\ref{sec:gate} with the same control Hamiltonians as in the previous experiments. We consider white noise processes $S_0=|0\rangle\langle 0|$ and $S_1=|1\rangle\langle 1|$, with $\gamma_0=0.14$ and $\gamma_1=0.07$ respectively. To quantify errors, we analyze the mean and variance of the quantity $J_\text{err}:=1-\Tr[U_T^\dagger V_T]$. Note that $S^\dagger_l S_l\neq I$, indicating that the fidelity $J_3$ may depend on the control Hamiltonian, as shown in Sec.~\ref{sec:gate}.

\begin{figure}
    \centering
    \includegraphics[width=\linewidth]{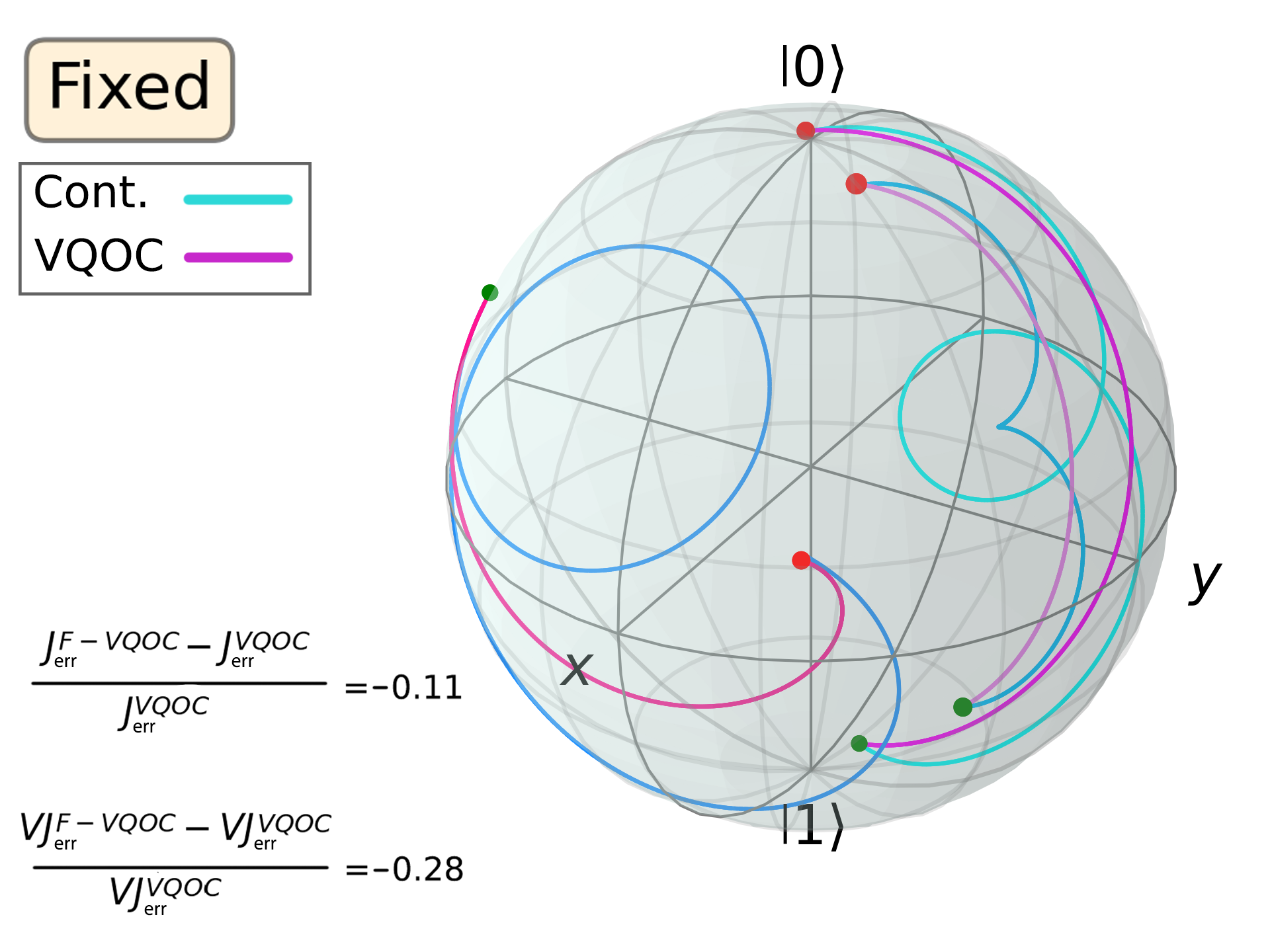}
    \caption{Example of gate optimization for a single random target unitary $U_{\text{targ}}$ done using F-VQOC vs.\ VQOC with $(\lambda,\mu,\nu)=(0.1,60,1)$ and $H_i=\sigma_i$, $S_0=|0\rangle\langle 0|$ and $S_1=|1\rangle\langle1|$ with white noise with $(\gamma_{0},\gamma_{1})=(0.14,0.07)$. Shown are three random initial states (green) and their paths taken by F-VQOC (blue lines) and VQOC (pink lines) to their target states (red).} 
    \label{fig:individualgate}
\end{figure}

Figure~\ref{fig:individualgate} illustrates one specific random target unitary demonstrating a significant decrease in both infidelity $J$ and variance $V\!J$ over all possible initial states of the gate. Interestingly, we observe that the paths tend to loop around a certain axis of the Bloch sphere, resembling trajectories that stay close to the noiseless eigenstates, as shown in Fig.~\ref{fig:onequbitfixedsingle}. 

\begin{figure}
    \centering
    \includegraphics[width=0.85\linewidth]{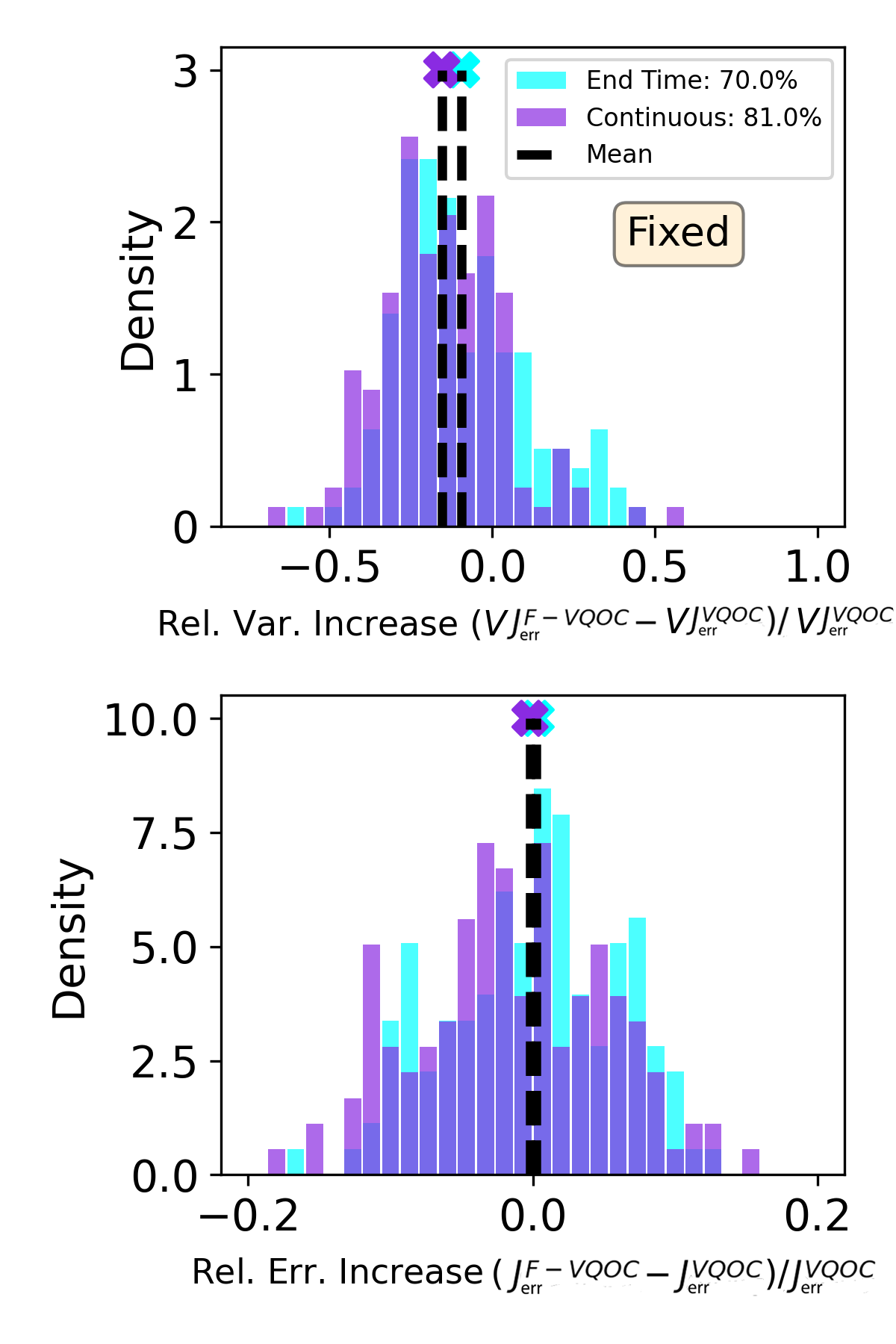}
    \caption{Relative unitary overlap error increase density plots  F-VQOC vs.\ VQOC with $(\lambda,\mu,\nu)=(0.1,60,1)$ and $H_i=\sigma_i$, $S_0=|0\rangle\langle 0|$ and $S_1=|1\rangle\langle1|$ with white noise with $(\gamma_{0},\gamma_{1})=(0.14,0.07)$ and 300 randomly sampled single-qubit unitaries $U_{\text{targ}}$ (one of which shown in detail in Fig.~\ref{fig:individualgate}). Legend indicates what percentage of F-VQOC trials outperform VQOC counterparts. Mean lines indicate mean relative error increase. a) relative variance increase, b) relative mean increase}
    \label{fig:gate}
\end{figure}

However, Fig.~\ref{fig:gate} shows that when sampling many possible target unitaries, the mean fidelities for VQOC and F-VQOC are similar. This is not surprising given the small gradients observed in the fidelity term $J_3$. Still, we do see different paths being taken through the Hilbert space (see Fig.~\ref{fig:individualgate}), highlighting the effect of the method, which is notable in the variances of the fidelities. F-VQOC shows a preference towards constructing pulses that result in similar fidelities for each input state, in contrast to having widely varying fidelity levels. 

\subsection{Multi-qubit}

To demonstrate the effectiveness of our methods for multi-qubit state preparation, we focus on generating Greenberger-Horne-Zeilinger (GHZ) states on two qubits, represented as $H_{\text{targ}}=|\text{GHZ}\rangle\langle \text{GHZ}|$. For each run, we use the parameters $(\lambda,\mu,\nu)=(0.1,600,1)$. The control Hamiltonian is given by $H=\prod_i \sigma_{Z,i}+\sum_i z_{X,i}\sigma_{X,i}+z_{Z,i}\sigma_{Z,i}$, where index $i$ corresponds to qubit $i$. For the noise operators, we consider $S_i=\sigma_{X,i}$ with $\gamma=0.07$, and $S_{ij}=\sigma_{Z,i}\sigma_{Z,j}$ with $\gamma=0.01$, accounting for noise at both single-qubit and multi-qubit levels.

Fig.~\ref{fig:multiqubit} shows that our method is capable of outperforming VQOC on a noisy multi-qubit system. As in the single qubit case, for general target states, the end time condition works well only with fixed noise, while the continuous case performs better in both fixed and scaled noise situations. This indicates that our method can effectively create control pulses for preparing important multi-qubit states, such as the GHZ state.\\

\section{Dicussion}\label{sec:conclusion}
This work introduces a versatile pulse optimization strategy based on the stochastic Schr\"{o}dinger equation. It offers insights into the fidelity landscape by utilizing the full distribution of states. Our gradient-based method, derived from stochastic optimal control, constructs high-fidelity pulses for a wide range of problems affected by various colored noise processes. This approach significantly reduces errors compared to noise-ignorant methods such as VQOC, particularly in ground-state preparation tasks. A key advantage of this algorithm is its ability to tailor pulses to the specific noise characteristics of a system. By inputting the noise operators and their strengths, the algorithm generates pulses that guide qubits through the least noise-sensitive areas of the Hilbert space. \\

In future research, our objective is to establish clearer relations between optimal regularization constants $(\lambda,\mu,\nu)$ and the system parameters. We also plan to expand our investigation to include arbitrary noise scaling. Additionally, we seek to gather more evidence supporting our hypothesis that continuous time costs outperform end-time costs in cases of linear noise scaling. We intend to experimentally validate our methods by comparing the fidelities of VQOC with F-VQOC pulses on a physical quantum computer. This could be achieved in an early-stage NISQ device by introducing artificial noise profiles to enhance the observed effects.

\begin{widetext2}
    \begin{minipage}[b]{\linewidth}
        \begin{figure}[H]
            \centering
            \includegraphics[scale=0.66]{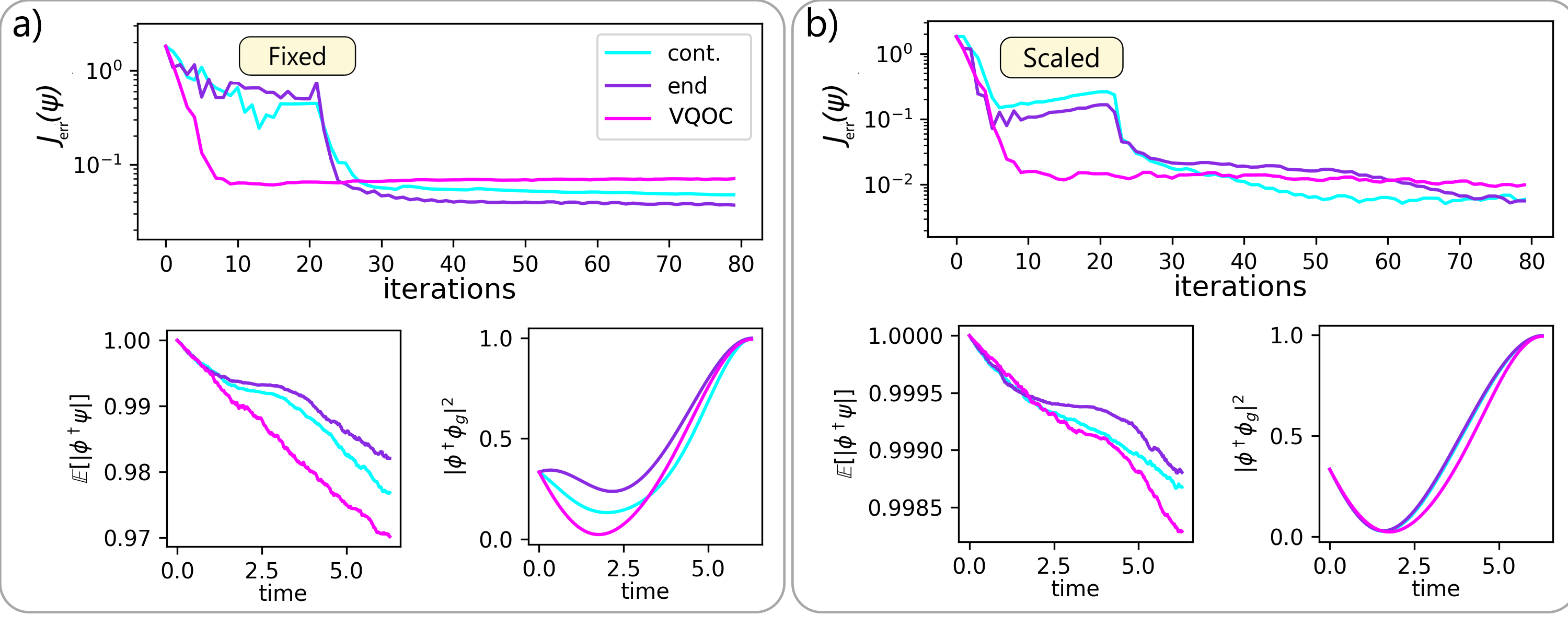}
    \caption{Average energy found per iteration for F-VQOC vs.\ VQOC with $(\lambda,\mu,\nu)=(0.1,600,1)$ and $H=\sum_i z_{X,i}\sigma_{X,i}+z_{Z,i}\sigma_{Z,i}$, $S_i=\sigma_{X,i}$ and $S_{ij}=\sigma_{Z,i}\sigma_{Z,j}$ with white noise with $\gamma_{X}=0.07$ and $\gamma_{ZZ}=0.01$. Below, fidelities plus overlap between the target state GHZ state and the noiseless state, indicating the path taken through state space. a) fixed noise strength, b) scaled noise strength.}
            \label{fig:multiqubit}
        \end{figure}    
    \end{minipage}
\end{widetext2}

\section*{ACKNOWLEDGEMENTS}
We thank Jasper Postema, Raul F.~Santos, Jasper van de Kraats, Madhav Mohan, and Emre Akaturk for fruitful discussions. This research is financially supported by the Dutch Ministry of Economic Affairs and Climate Policy (EZK), as part of the Quantum Delta NL program,  the Horizon Europe programme HORIZON-CL4-2021-DIGITAL-EMERGING-01-30 via the project 101070144 (EuRyQa), and by the Netherlands Organisation for Scientific Research (NWO) under Grant No.\ 680.92.18.05. L.Y.~Visser and O.~Tse acknowledge support from NWO grant NGF.1582.22.009.
\\

\section*{COMPETING INTERESTS}
The authors declare no competing interests.

\section*{DATA AVAILABILITY}
The data and code that support the findings of this study are available from the corresponding author upon reasonable request.

\newpage
\bibliography{Bibliography.bib}

\begin{thebibliography}{57}%
\makeatletter
\providecommand \@ifxundefined [1]{%
 \@ifx{#1\undefined}
}%
\providecommand \@ifnum [1]{%
 \ifnum #1\expandafter \@firstoftwo
 \else \expandafter \@secondoftwo
 \fi
}%
\providecommand \@ifx [1]{%
 \ifx #1\expandafter \@firstoftwo
 \else \expandafter \@secondoftwo
 \fi
}%
\providecommand \natexlab [1]{#1}%
\providecommand \enquote  [1]{``#1''}%
\providecommand \bibnamefont  [1]{#1}%
\providecommand \bibfnamefont [1]{#1}%
\providecommand \citenamefont [1]{#1}%
\providecommand \href@noop [0]{\@secondoftwo}%
\providecommand \href [0]{\begingroup \@sanitize@url \@href}%
\providecommand \@href[1]{\@@startlink{#1}\@@href}%
\providecommand \@@href[1]{\endgroup#1\@@endlink}%
\providecommand \@sanitize@url [0]{\catcode `\\12\catcode `\$12\catcode `\&12\catcode `\#12\catcode `\^12\catcode `\_12\catcode `\%12\relax}%
\providecommand \@@startlink[1]{}%
\providecommand \@@endlink[0]{}%
\providecommand \url  [0]{\begingroup\@sanitize@url \@url }%
\providecommand \@url [1]{\endgroup\@href {#1}{\urlprefix }}%
\providecommand \urlprefix  [0]{URL }%
\providecommand \Eprint [0]{\href }%
\providecommand \doibase [0]{https://doi.org/}%
\providecommand \selectlanguage [0]{\@gobble}%
\providecommand \bibinfo  [0]{\@secondoftwo}%
\providecommand \bibfield  [0]{\@secondoftwo}%
\providecommand \translation [1]{[#1]}%
\providecommand \BibitemOpen [0]{}%
\providecommand \bibitemStop [0]{}%
\providecommand \bibitemNoStop [0]{.\EOS\space}%
\providecommand \EOS [0]{\spacefactor3000\relax}%
\providecommand \BibitemShut  [1]{\csname bibitem#1\endcsname}%
\let\auto@bib@innerbib\@empty
\bibitem [{\citenamefont {Preskill}(2018)}]{Preskill_2018}%
  \BibitemOpen
  \bibfield  {author} {\bibinfo {author} {\bibfnamefont {J.}~\bibnamefont {Preskill}},\ }\bibfield  {title} {\bibinfo {title} {Quantum computing in the nisq era and beyond},\ }\href {https://doi.org/10.22331/q-2018-08-06-79} {\bibfield  {journal} {\bibinfo  {journal} {Quantum}\ }\textbf {\bibinfo {volume} {2}},\ \bibinfo {pages} {79} (\bibinfo {year} {2018})}\BibitemShut {NoStop}%
\bibitem [{\citenamefont {Morgado}\ and\ \citenamefont {Whitlock}(2021)}]{morgado}%
  \BibitemOpen
  \bibfield  {author} {\bibinfo {author} {\bibfnamefont {M.}~\bibnamefont {Morgado}}\ and\ \bibinfo {author} {\bibfnamefont {S.}~\bibnamefont {Whitlock}},\ }\bibfield  {title} {\bibinfo {title} {Quantum simulation and computing with rydberg-interacting qubits},\ }\href {https://doi.org/10.1116/5.0036562} {\bibfield  {journal} {\bibinfo  {journal} {AVS Quantum Science}\ }\textbf {\bibinfo {volume} {3}},\ \bibinfo {pages} {023501} (\bibinfo {year} {2021})},\ \Eprint {https://arxiv.org/abs/https://pubs.aip.org/avs/aqs/article-pdf/doi/10.1116/5.0036562/19739152/023501\_1\_online.pdf} {https://pubs.aip.org/avs/aqs/article-pdf/doi/10.1116/5.0036562/19739152/023501\_1\_online.pdf} \BibitemShut {NoStop}%
\bibitem [{\citenamefont {Aseguinolaza}\ \emph {et~al.}(2024)\citenamefont {Aseguinolaza}, \citenamefont {Sobrino}, \citenamefont {Sobrino}, \citenamefont {Jornet-Somoza},\ and\ \citenamefont {Borge}}]{errors1}%
  \BibitemOpen
  \bibfield  {author} {\bibinfo {author} {\bibfnamefont {U.}~\bibnamefont {Aseguinolaza}}, \bibinfo {author} {\bibfnamefont {N.}~\bibnamefont {Sobrino}}, \bibinfo {author} {\bibfnamefont {G.}~\bibnamefont {Sobrino}}, \bibinfo {author} {\bibfnamefont {J.}~\bibnamefont {Jornet-Somoza}},\ and\ \bibinfo {author} {\bibfnamefont {J.}~\bibnamefont {Borge}},\ }\bibfield  {title} {\bibinfo {title} {Error estimation in current noisy quantum computers},\ }\href {https://doi.org/10.1007/s11128-024-04384-z} {\bibfield  {journal} {\bibinfo  {journal} {Quantum Information Processing}\ }\textbf {\bibinfo {volume} {23}},\ \bibinfo {pages} {181} (\bibinfo {year} {2024})}\BibitemShut {NoStop}%
\bibitem [{\citenamefont {Pagano}\ \emph {et~al.}(2022)\citenamefont {Pagano}, \citenamefont {Weber}, \citenamefont {Jaschke}, \citenamefont {Pfau}, \citenamefont {Meinert}, \citenamefont {Montangero},\ and\ \citenamefont {B\"uchler}}]{errors2}%
  \BibitemOpen
  \bibfield  {author} {\bibinfo {author} {\bibfnamefont {A.}~\bibnamefont {Pagano}}, \bibinfo {author} {\bibfnamefont {S.}~\bibnamefont {Weber}}, \bibinfo {author} {\bibfnamefont {D.}~\bibnamefont {Jaschke}}, \bibinfo {author} {\bibfnamefont {T.}~\bibnamefont {Pfau}}, \bibinfo {author} {\bibfnamefont {F.}~\bibnamefont {Meinert}}, \bibinfo {author} {\bibfnamefont {S.}~\bibnamefont {Montangero}},\ and\ \bibinfo {author} {\bibfnamefont {H.~P.}\ \bibnamefont {B\"uchler}},\ }\bibfield  {title} {\bibinfo {title} {Error budgeting for a controlled-phase gate with strontium-88 rydberg atoms},\ }\href {https://doi.org/10.1103/PhysRevResearch.4.033019} {\bibfield  {journal} {\bibinfo  {journal} {Phys. Rev. Res.}\ }\textbf {\bibinfo {volume} {4}},\ \bibinfo {pages} {033019} (\bibinfo {year} {2022})}\BibitemShut {NoStop}%
\bibitem [{\citenamefont {Wiening}\ \emph {et~al.}(2024)\citenamefont {Wiening}, \citenamefont {Bergendahl}, \citenamefont {Leyton-Ortega},\ and\ \citenamefont {Nalbach}}]{statepreppulse}%
  \BibitemOpen
  \bibfield  {author} {\bibinfo {author} {\bibfnamefont {A.~S.}\ \bibnamefont {Wiening}}, \bibinfo {author} {\bibfnamefont {J.}~\bibnamefont {Bergendahl}}, \bibinfo {author} {\bibfnamefont {V.}~\bibnamefont {Leyton-Ortega}},\ and\ \bibinfo {author} {\bibfnamefont {P.}~\bibnamefont {Nalbach}},\ }\href {https://arxiv.org/abs/2409.08204} {\bibinfo {title} {Optimizing qubit control pulses for state preparation}} (\bibinfo {year} {2024}),\ \Eprint {https://arxiv.org/abs/2409.08204} {arXiv:2409.08204 [quant-ph]} \BibitemShut {NoStop}%
\bibitem [{\citenamefont {Meitei}\ \emph {et~al.}(2021)\citenamefont {Meitei}, \citenamefont {Gard}, \citenamefont {Barron}, \citenamefont {Pappas}, \citenamefont {Economou}, \citenamefont {Barnes},\ and\ \citenamefont {Mayhall}}]{qocvqe1}%
  \BibitemOpen
  \bibfield  {author} {\bibinfo {author} {\bibfnamefont {O.~R.}\ \bibnamefont {Meitei}}, \bibinfo {author} {\bibfnamefont {B.~T.}\ \bibnamefont {Gard}}, \bibinfo {author} {\bibfnamefont {G.~S.}\ \bibnamefont {Barron}}, \bibinfo {author} {\bibfnamefont {D.~P.}\ \bibnamefont {Pappas}}, \bibinfo {author} {\bibfnamefont {S.~E.}\ \bibnamefont {Economou}}, \bibinfo {author} {\bibfnamefont {E.}~\bibnamefont {Barnes}},\ and\ \bibinfo {author} {\bibfnamefont {N.~J.}\ \bibnamefont {Mayhall}},\ }\bibfield  {title} {\bibinfo {title} {Gate-free state preparation for fast variational quantum eigensolver simulations},\ }\href {https://doi.org/10.1038/s41534-021-00493-0} {\bibfield  {journal} {\bibinfo  {journal} {npj Quantum Information}\ }\textbf {\bibinfo {volume} {7}},\ \bibinfo {pages} {155} (\bibinfo {year} {2021})}\BibitemShut {NoStop}%
\bibitem [{\citenamefont {Choquette}\ \emph {et~al.}(2021)\citenamefont {Choquette}, \citenamefont {Di~Paolo}, \citenamefont {Barkoutsos}, \citenamefont {S\'en\'echal}, \citenamefont {Tavernelli},\ and\ \citenamefont {Blais}}]{qocvqe2}%
  \BibitemOpen
  \bibfield  {author} {\bibinfo {author} {\bibfnamefont {A.}~\bibnamefont {Choquette}}, \bibinfo {author} {\bibfnamefont {A.}~\bibnamefont {Di~Paolo}}, \bibinfo {author} {\bibfnamefont {P.~K.}\ \bibnamefont {Barkoutsos}}, \bibinfo {author} {\bibfnamefont {D.}~\bibnamefont {S\'en\'echal}}, \bibinfo {author} {\bibfnamefont {I.}~\bibnamefont {Tavernelli}},\ and\ \bibinfo {author} {\bibfnamefont {A.}~\bibnamefont {Blais}},\ }\bibfield  {title} {\bibinfo {title} {Quantum-optimal-control-inspired ansatz for variational quantum algorithms},\ }\href {https://doi.org/10.1103/PhysRevResearch.3.023092} {\bibfield  {journal} {\bibinfo  {journal} {Phys. Rev. Research}\ }\textbf {\bibinfo {volume} {3}},\ \bibinfo {pages} {023092} (\bibinfo {year} {2021})}\BibitemShut {NoStop}%
\bibitem [{\citenamefont {Jandura}\ and\ \citenamefont {Pupillo}(2022)}]{jandura}%
  \BibitemOpen
  \bibfield  {author} {\bibinfo {author} {\bibfnamefont {S.}~\bibnamefont {Jandura}}\ and\ \bibinfo {author} {\bibfnamefont {G.}~\bibnamefont {Pupillo}},\ }\bibfield  {title} {\bibinfo {title} {Time-{O}ptimal {T}wo- and {T}hree-{Q}ubit {G}ates for {R}ydberg {A}toms},\ }\href {https://doi.org/10.22331/q-2022-05-13-712} {\bibfield  {journal} {\bibinfo  {journal} {{Quantum}}\ }\textbf {\bibinfo {volume} {6}},\ \bibinfo {pages} {712} (\bibinfo {year} {2022})}\BibitemShut {NoStop}%
\bibitem [{\citenamefont {Dionis}\ and\ \citenamefont {Sugny}(2023)}]{timeoptimal2}%
  \BibitemOpen
  \bibfield  {author} {\bibinfo {author} {\bibfnamefont {E.}~\bibnamefont {Dionis}}\ and\ \bibinfo {author} {\bibfnamefont {D.}~\bibnamefont {Sugny}},\ }\bibfield  {title} {\bibinfo {title} {Time-optimal control of two-level quantum systems by piecewise constant pulses},\ }\href {https://doi.org/10.1103/PhysRevA.107.032613} {\bibfield  {journal} {\bibinfo  {journal} {Phys. Rev. A}\ }\textbf {\bibinfo {volume} {107}},\ \bibinfo {pages} {032613} (\bibinfo {year} {2023})}\BibitemShut {NoStop}%
\bibitem [{\citenamefont {Gupta}\ and\ \citenamefont {Tembulkar}(1984)}]{dd1}%
  \BibitemOpen
  \bibfield  {author} {\bibinfo {author} {\bibfnamefont {A.~K.}\ \bibnamefont {Gupta}}\ and\ \bibinfo {author} {\bibfnamefont {J.~M.}\ \bibnamefont {Tembulkar}},\ }\bibfield  {title} {\bibinfo {title} {Dynamic decoupling of secondary systems},\ }\href {https://doi.org/https://doi.org/10.1016/0029-5493(84)90282-6} {\bibfield  {journal} {\bibinfo  {journal} {Nuclear Engineering and Design}\ }\textbf {\bibinfo {volume} {81}},\ \bibinfo {pages} {359} (\bibinfo {year} {1984})}\BibitemShut {NoStop}%
\bibitem [{\citenamefont {Ezzell}\ \emph {et~al.}(2023)\citenamefont {Ezzell}, \citenamefont {Pokharel}, \citenamefont {Tewala}, \citenamefont {Quiroz},\ and\ \citenamefont {Lidar}}]{dd2}%
  \BibitemOpen
  \bibfield  {author} {\bibinfo {author} {\bibfnamefont {N.}~\bibnamefont {Ezzell}}, \bibinfo {author} {\bibfnamefont {B.}~\bibnamefont {Pokharel}}, \bibinfo {author} {\bibfnamefont {L.}~\bibnamefont {Tewala}}, \bibinfo {author} {\bibfnamefont {G.}~\bibnamefont {Quiroz}},\ and\ \bibinfo {author} {\bibfnamefont {D.~A.}\ \bibnamefont {Lidar}},\ }\bibfield  {title} {\bibinfo {title} {Dynamical decoupling for superconducting qubits: A performance survey},\ }\href {https://doi.org/10.1103/PhysRevApplied.20.064027} {\bibfield  {journal} {\bibinfo  {journal} {Phys. Rev. Appl.}\ }\textbf {\bibinfo {volume} {20}},\ \bibinfo {pages} {064027} (\bibinfo {year} {2023})}\BibitemShut {NoStop}%
\bibitem [{\citenamefont {Mohan}\ \emph {et~al.}(2023)\citenamefont {Mohan}, \citenamefont {de~Keijzer},\ and\ \citenamefont {Kokkelmans}}]{madhav}%
  \BibitemOpen
  \bibfield  {author} {\bibinfo {author} {\bibfnamefont {M.}~\bibnamefont {Mohan}}, \bibinfo {author} {\bibfnamefont {R.}~\bibnamefont {de~Keijzer}},\ and\ \bibinfo {author} {\bibfnamefont {S.}~\bibnamefont {Kokkelmans}},\ }\bibfield  {title} {\bibinfo {title} {Robust control and optimal rydberg states for neutral atom two-qubit gates},\ }\href {https://doi.org/10.1103/PhysRevResearch.5.033052} {\bibfield  {journal} {\bibinfo  {journal} {Phys. Rev. Res.}\ }\textbf {\bibinfo {volume} {5}},\ \bibinfo {pages} {033052} (\bibinfo {year} {2023})}\BibitemShut {NoStop}%
\bibitem [{\citenamefont {Kosut}\ \emph {et~al.}(2022)\citenamefont {Kosut}, \citenamefont {Bhole},\ and\ \citenamefont {Rabitz}}]{kosut2022robust}%
  \BibitemOpen
  \bibfield  {author} {\bibinfo {author} {\bibfnamefont {R.~L.}\ \bibnamefont {Kosut}}, \bibinfo {author} {\bibfnamefont {G.}~\bibnamefont {Bhole}},\ and\ \bibinfo {author} {\bibfnamefont {H.}~\bibnamefont {Rabitz}},\ }\href@noop {} {\bibinfo {title} {Robust quantum control: Analysis \& synthesis via averaging}} (\bibinfo {year} {2022}),\ \Eprint {https://arxiv.org/abs/2208.14193} {arXiv:2208.14193 [quant-ph]} \BibitemShut {NoStop}%
\bibitem [{\citenamefont {Shi}\ \emph {et~al.}(2024)\citenamefont {Shi}, \citenamefont {Ding}, \citenamefont {Chen}, \citenamefont {Song}, \citenamefont {Xia}, \citenamefont {Yi},\ and\ \citenamefont {Nori}}]{robust2}%
  \BibitemOpen
  \bibfield  {author} {\bibinfo {author} {\bibfnamefont {Z.-C.}\ \bibnamefont {Shi}}, \bibinfo {author} {\bibfnamefont {J.-T.}\ \bibnamefont {Ding}}, \bibinfo {author} {\bibfnamefont {Y.-H.}\ \bibnamefont {Chen}}, \bibinfo {author} {\bibfnamefont {J.}~\bibnamefont {Song}}, \bibinfo {author} {\bibfnamefont {Y.}~\bibnamefont {Xia}}, \bibinfo {author} {\bibfnamefont {X.}~\bibnamefont {Yi}},\ and\ \bibinfo {author} {\bibfnamefont {F.}~\bibnamefont {Nori}},\ }\bibfield  {title} {\bibinfo {title} {Supervised learning for robust quantum control in composite-pulse systems},\ }\href {https://doi.org/10.1103/PhysRevApplied.21.044012} {\bibfield  {journal} {\bibinfo  {journal} {Phys. Rev. Appl.}\ }\textbf {\bibinfo {volume} {21}},\ \bibinfo {pages} {044012} (\bibinfo {year} {2024})}\BibitemShut {NoStop}%
\bibitem [{\citenamefont {jun Zhu}\ \emph {et~al.}(2022)\citenamefont {jun Zhu}, \citenamefont {Laforgue}, \citenamefont {Chen},\ and\ \citenamefont {Guérin}}]{robust3}%
  \BibitemOpen
  \bibfield  {author} {\bibinfo {author} {\bibfnamefont {J.}~\bibnamefont {jun Zhu}}, \bibinfo {author} {\bibfnamefont {X.}~\bibnamefont {Laforgue}}, \bibinfo {author} {\bibfnamefont {X.}~\bibnamefont {Chen}},\ and\ \bibinfo {author} {\bibfnamefont {S.}~\bibnamefont {Guérin}},\ }\bibfield  {title} {\bibinfo {title} {Robust quantum control by smooth quasi-square pulses},\ }\href {https://doi.org/10.1088/1361-6455/ac8adf} {\bibfield  {journal} {\bibinfo  {journal} {Journal of Physics B: Atomic, Molecular and Optical Physics}\ }\textbf {\bibinfo {volume} {55}},\ \bibinfo {pages} {194001} (\bibinfo {year} {2022})}\BibitemShut {NoStop}%
\bibitem [{\citenamefont {Poggi}\ \emph {et~al.}(2023)\citenamefont {Poggi}, \citenamefont {Chiara}, \citenamefont {Campbell},\ and\ \citenamefont {Kiely}}]{Poggi2023UniversallyRQ}%
  \BibitemOpen
  \bibfield  {author} {\bibinfo {author} {\bibfnamefont {P.~M.}\ \bibnamefont {Poggi}}, \bibinfo {author} {\bibfnamefont {G.~D.}\ \bibnamefont {Chiara}}, \bibinfo {author} {\bibfnamefont {S.}~\bibnamefont {Campbell}},\ and\ \bibinfo {author} {\bibfnamefont {A.}~\bibnamefont {Kiely}},\ }\bibfield  {title} {\bibinfo {title} {Universally robust quantum control.},\ }\href {https://api.semanticscholar.org/CorpusID:262822629} {\bibfield  {journal} {\bibinfo  {journal} {Physical review letters}\ }\textbf {\bibinfo {volume} {132 19}},\ \bibinfo {pages} {193801} (\bibinfo {year} {2023})}\BibitemShut {NoStop}%
\bibitem [{\citenamefont {Weidner}\ \emph {et~al.}(2023)\citenamefont {Weidner}, \citenamefont {Reed}, \citenamefont {Monroe}, \citenamefont {Sheller}, \citenamefont {O'Neil}, \citenamefont {Maas}, \citenamefont {Jonckheere}, \citenamefont {Langbein},\ and\ \citenamefont {Schirmer}}]{weidner2023robust}%
  \BibitemOpen
  \bibfield  {author} {\bibinfo {author} {\bibfnamefont {C.~A.}\ \bibnamefont {Weidner}}, \bibinfo {author} {\bibfnamefont {E.~A.}\ \bibnamefont {Reed}}, \bibinfo {author} {\bibfnamefont {J.}~\bibnamefont {Monroe}}, \bibinfo {author} {\bibfnamefont {B.}~\bibnamefont {Sheller}}, \bibinfo {author} {\bibfnamefont {S.}~\bibnamefont {O'Neil}}, \bibinfo {author} {\bibfnamefont {E.}~\bibnamefont {Maas}}, \bibinfo {author} {\bibfnamefont {E.~A.}\ \bibnamefont {Jonckheere}}, \bibinfo {author} {\bibfnamefont {F.~C.}\ \bibnamefont {Langbein}},\ and\ \bibinfo {author} {\bibfnamefont {S.~G.}\ \bibnamefont {Schirmer}},\ }\href@noop {} {\bibinfo {title} {Robust quantum control in closed and open systems: Theory and practice}} (\bibinfo {year} {2023}),\ \Eprint {https://arxiv.org/abs/2401.00294} {arXiv:2401.00294 [quant-ph]} \BibitemShut {NoStop}%
\bibitem [{\citenamefont {Aroch}\ \emph {et~al.}(2023)\citenamefont {Aroch}, \citenamefont {Kosloff},\ and\ \citenamefont {Kallush}}]{aroch2023mitigating}%
  \BibitemOpen
  \bibfield  {author} {\bibinfo {author} {\bibfnamefont {A.}~\bibnamefont {Aroch}}, \bibinfo {author} {\bibfnamefont {R.}~\bibnamefont {Kosloff}},\ and\ \bibinfo {author} {\bibfnamefont {S.}~\bibnamefont {Kallush}},\ }\href@noop {} {\bibinfo {title} {Mitigating controller noise in quantum gates using optimal control theory}} (\bibinfo {year} {2023}),\ \Eprint {https://arxiv.org/abs/2309.07659} {arXiv:2309.07659 [quant-ph]} \BibitemShut {NoStop}%
\bibitem [{\citenamefont {Lin}\ \emph {et~al.}(2020)\citenamefont {Lin}, \citenamefont {Sels}, \citenamefont {Ma},\ and\ \citenamefont {Wang}}]{PMPquantumcontrol}%
  \BibitemOpen
  \bibfield  {author} {\bibinfo {author} {\bibfnamefont {C.}~\bibnamefont {Lin}}, \bibinfo {author} {\bibfnamefont {D.}~\bibnamefont {Sels}}, \bibinfo {author} {\bibfnamefont {Y.}~\bibnamefont {Ma}},\ and\ \bibinfo {author} {\bibfnamefont {Y.}~\bibnamefont {Wang}},\ }\bibfield  {title} {\bibinfo {title} {Stochastic optimal control formalism for an open quantum system},\ }\href {https://doi.org/10.1103/PhysRevA.102.052605} {\bibfield  {journal} {\bibinfo  {journal} {Phys. Rev. A}\ }\textbf {\bibinfo {volume} {102}},\ \bibinfo {pages} {052605} (\bibinfo {year} {2020})}\BibitemShut {NoStop}%
\bibitem [{\citenamefont {Vinter}(2013)}]{pontryagin}%
  \BibitemOpen
  \bibfield  {author} {\bibinfo {author} {\bibfnamefont {R.~B.}\ \bibnamefont {Vinter}},\ }\bibinfo {title} {Optimal control and pontryagin's maximum principle},\ in\ \href {https://doi.org/10.1007/978-1-4471-5102-9_200-1} {\emph {\bibinfo {booktitle} {Encyclopedia of Systems and Control}}},\ \bibinfo {editor} {edited by\ \bibinfo {editor} {\bibfnamefont {J.}~\bibnamefont {Baillieul}}\ and\ \bibinfo {editor} {\bibfnamefont {T.}~\bibnamefont {Samad}}}\ (\bibinfo  {publisher} {Springer London},\ \bibinfo {address} {London},\ \bibinfo {year} {2013})\ pp.\ \bibinfo {pages} {1--9}\BibitemShut {NoStop}%
\bibitem [{\citenamefont {Moodley}\ and\ \citenamefont {Petruccione}(2009)}]{unravelling}%
  \BibitemOpen
  \bibfield  {author} {\bibinfo {author} {\bibfnamefont {M.}~\bibnamefont {Moodley}}\ and\ \bibinfo {author} {\bibfnamefont {F.}~\bibnamefont {Petruccione}},\ }\bibfield  {title} {\bibinfo {title} {Stochastic wave-function unraveling of the generalized lindblad master equation},\ }\href {https://doi.org/10.1103/PhysRevA.79.042103} {\bibfield  {journal} {\bibinfo  {journal} {Phys. Rev. A}\ }\textbf {\bibinfo {volume} {79}},\ \bibinfo {pages} {042103} (\bibinfo {year} {2009})}\BibitemShut {NoStop}%
\bibitem [{\citenamefont {Barchielli}\ \emph {et~al.}(2010)\citenamefont {Barchielli}, \citenamefont {Pellegrini},\ and\ \citenamefont {Petruccione}}]{Barchielli_2010}%
  \BibitemOpen
  \bibfield  {author} {\bibinfo {author} {\bibfnamefont {A.}~\bibnamefont {Barchielli}}, \bibinfo {author} {\bibfnamefont {C.}~\bibnamefont {Pellegrini}},\ and\ \bibinfo {author} {\bibfnamefont {F.}~\bibnamefont {Petruccione}},\ }\bibfield  {title} {\bibinfo {title} {Stochastic schr\"odinger equations with coloured noise},\ }\href {https://doi.org/10.1209/0295-5075/91/24001} {\bibfield  {journal} {\bibinfo  {journal} {Europhysics Letters}\ }\textbf {\bibinfo {volume} {91}},\ \bibinfo {pages} {24001} (\bibinfo {year} {2010})}\BibitemShut {NoStop}%
\bibitem [{\citenamefont {Gardiner}\ and\ \citenamefont {Zoller}(2004)}]{gardiner2004quantum}%
  \BibitemOpen
  \bibfield  {author} {\bibinfo {author} {\bibfnamefont {C.}~\bibnamefont {Gardiner}}\ and\ \bibinfo {author} {\bibfnamefont {P.}~\bibnamefont {Zoller}},\ }\href {https://books.google.nl/books?id=a_xsT8oGhdgC} {\emph {\bibinfo {title} {Quantum Noise: A Handbook of Markovian and Non-Markovian Quantum Stochastic Methods with Applications to Quantum Optics}}},\ Springer Series in Synergetics\ (\bibinfo  {publisher} {Springer},\ \bibinfo {year} {2004})\BibitemShut {NoStop}%
\bibitem [{\citenamefont {de~Keijzer}\ \emph {et~al.}(2024)\citenamefont {de~Keijzer}, \citenamefont {Visser}, \citenamefont {Tse},\ and\ \citenamefont {Kokkelmans}}]{ssepaper}%
  \BibitemOpen
  \bibfield  {author} {\bibinfo {author} {\bibfnamefont {R.}~\bibnamefont {de~Keijzer}}, \bibinfo {author} {\bibfnamefont {L.}~\bibnamefont {Visser}}, \bibinfo {author} {\bibfnamefont {O.}~\bibnamefont {Tse}},\ and\ \bibinfo {author} {\bibfnamefont {S.}~\bibnamefont {Kokkelmans}},\ }\href@noop {} {\bibinfo {title} {Qubit fidelity under stochastic schr\"odinger equations driven by colored noise}} (\bibinfo {year} {2024}),\ \Eprint {https://arxiv.org/abs/2401.11758} {arXiv:2401.11758 [quant-ph]} \BibitemShut {NoStop}%
\bibitem [{\citenamefont {Day}\ \emph {et~al.}(2022)\citenamefont {Day}, \citenamefont {Low}, \citenamefont {White}, \citenamefont {Islam},\ and\ \citenamefont {Senko}}]{parameters2}%
  \BibitemOpen
  \bibfield  {author} {\bibinfo {author} {\bibfnamefont {M.~L.}\ \bibnamefont {Day}}, \bibinfo {author} {\bibfnamefont {P.~J.}\ \bibnamefont {Low}}, \bibinfo {author} {\bibfnamefont {B.}~\bibnamefont {White}}, \bibinfo {author} {\bibfnamefont {R.}~\bibnamefont {Islam}},\ and\ \bibinfo {author} {\bibfnamefont {C.}~\bibnamefont {Senko}},\ }\bibfield  {title} {\bibinfo {title} {Limits on atomic qubit control from laser noise},\ }\href {https://doi.org/10.1038/s41534-022-00586-4} {\bibfield  {journal} {\bibinfo  {journal} {npj Quantum Information}\ }\textbf {\bibinfo {volume} {8}},\ \bibinfo {pages} {72} (\bibinfo {year} {2022})}\BibitemShut {NoStop}%
\bibitem [{\citenamefont {Rower}\ \emph {et~al.}(2023)\citenamefont {Rower}, \citenamefont {Ateshian}, \citenamefont {Li}, \citenamefont {Hays}, \citenamefont {Bluvstein}, \citenamefont {Ding}, \citenamefont {Kannan}, \citenamefont {Almanakly}, \citenamefont {Braum\"uller}, \citenamefont {Kim}, \citenamefont {Melville}, \citenamefont {Niedzielski}, \citenamefont {Schwartz}, \citenamefont {Yoder}, \citenamefont {Orlando}, \citenamefont {Wang}, \citenamefont {Gustavsson}, \citenamefont {Grover}, \citenamefont {Serniak}, \citenamefont {Comin},\ and\ \citenamefont {Oliver}}]{fluxnoise}%
  \BibitemOpen
  \bibfield  {author} {\bibinfo {author} {\bibfnamefont {D.~A.}\ \bibnamefont {Rower}}, \bibinfo {author} {\bibfnamefont {L.}~\bibnamefont {Ateshian}}, \bibinfo {author} {\bibfnamefont {L.~H.}\ \bibnamefont {Li}}, \bibinfo {author} {\bibfnamefont {M.}~\bibnamefont {Hays}}, \bibinfo {author} {\bibfnamefont {D.}~\bibnamefont {Bluvstein}}, \bibinfo {author} {\bibfnamefont {L.}~\bibnamefont {Ding}}, \bibinfo {author} {\bibfnamefont {B.}~\bibnamefont {Kannan}}, \bibinfo {author} {\bibfnamefont {A.}~\bibnamefont {Almanakly}}, \bibinfo {author} {\bibfnamefont {J.}~\bibnamefont {Braum\"uller}}, \bibinfo {author} {\bibfnamefont {D.~K.}\ \bibnamefont {Kim}}, \bibinfo {author} {\bibfnamefont {A.}~\bibnamefont {Melville}}, \bibinfo {author} {\bibfnamefont {B.~M.}\ \bibnamefont {Niedzielski}}, \bibinfo {author} {\bibfnamefont {M.~E.}\ \bibnamefont {Schwartz}}, \bibinfo {author} {\bibfnamefont {J.~L.}\ \bibnamefont {Yoder}}, \bibinfo {author} {\bibfnamefont {T.~P.}\ \bibnamefont {Orlando}}, \bibinfo {author} {\bibfnamefont
  {J.~I.-J.}\ \bibnamefont {Wang}}, \bibinfo {author} {\bibfnamefont {S.}~\bibnamefont {Gustavsson}}, \bibinfo {author} {\bibfnamefont {J.~A.}\ \bibnamefont {Grover}}, \bibinfo {author} {\bibfnamefont {K.}~\bibnamefont {Serniak}}, \bibinfo {author} {\bibfnamefont {R.}~\bibnamefont {Comin}},\ and\ \bibinfo {author} {\bibfnamefont {W.~D.}\ \bibnamefont {Oliver}},\ }\bibfield  {title} {\bibinfo {title} {Evolution of $1/f$ flux noise in superconducting qubits with weak magnetic fields},\ }\href {https://doi.org/10.1103/PhysRevLett.130.220602} {\bibfield  {journal} {\bibinfo  {journal} {Phys. Rev. Lett.}\ }\textbf {\bibinfo {volume} {130}},\ \bibinfo {pages} {220602} (\bibinfo {year} {2023})}\BibitemShut {NoStop}%
\bibitem [{\citenamefont {Semina}\ \emph {et~al.}(2014)\citenamefont {Semina}, \citenamefont {Semin}, \citenamefont {Petruccione},\ and\ \citenamefont {Barchielli}}]{semina}%
  \BibitemOpen
  \bibfield  {author} {\bibinfo {author} {\bibfnamefont {I.}~\bibnamefont {Semina}}, \bibinfo {author} {\bibfnamefont {V.}~\bibnamefont {Semin}}, \bibinfo {author} {\bibfnamefont {F.}~\bibnamefont {Petruccione}},\ and\ \bibinfo {author} {\bibfnamefont {A.}~\bibnamefont {Barchielli}},\ }\bibfield  {title} {\bibinfo {title} {Stochastic schrödinger equations for markovian and non-markovian cases},\ }\href {https://doi.org/10.1142/S1230161214400083} {\bibfield  {journal} {\bibinfo  {journal} {Open Systems \& Information Dynamics}\ }\textbf {\bibinfo {volume} {21}},\ \bibinfo {pages} {1440008} (\bibinfo {year} {2014})}\BibitemShut {NoStop}%
\bibitem [{\citenamefont {Breuer}\ and\ \citenamefont {Piilo}(2009)}]{unraveltome}%
  \BibitemOpen
  \bibfield  {author} {\bibinfo {author} {\bibfnamefont {H.-P.}\ \bibnamefont {Breuer}}\ and\ \bibinfo {author} {\bibfnamefont {J.}~\bibnamefont {Piilo}},\ }\bibfield  {title} {\bibinfo {title} {Stochastic jump processes for non-markovian quantum dynamics},\ }\href {https://doi.org/10.1209/0295-5075/85/50004} {\bibfield  {journal} {\bibinfo  {journal} {EPL (Europhysics Letters)}\ }\textbf {\bibinfo {volume} {85}},\ \bibinfo {pages} {50004} (\bibinfo {year} {2009})}\BibitemShut {NoStop}%
\bibitem [{\citenamefont {Cui}\ \emph {et~al.}(2023)\citenamefont {Cui}, \citenamefont {Liu},\ and\ \citenamefont {Zhou}}]{wavefunctioncontrol}%
  \BibitemOpen
  \bibfield  {author} {\bibinfo {author} {\bibfnamefont {J.}~\bibnamefont {Cui}}, \bibinfo {author} {\bibfnamefont {S.}~\bibnamefont {Liu}},\ and\ \bibinfo {author} {\bibfnamefont {H.}~\bibnamefont {Zhou}},\ }\bibfield  {title} {\bibinfo {title} {Optimal control for stochastic nonlinear schrödinger equation on graph},\ }\href {https://doi.org/10.1137/22M1524175} {\bibfield  {journal} {\bibinfo  {journal} {SIAM Journal on Control and Optimization}\ }\textbf {\bibinfo {volume} {61}},\ \bibinfo {pages} {2021} (\bibinfo {year} {2023})}\BibitemShut {NoStop}%
\bibitem [{\citenamefont {Breckner}\ \emph {et~al.}(2021)\citenamefont {Breckner}, \citenamefont {Lisei},\ and\ \citenamefont {{Ionuţ Şimon}}}]{mathematicalsse}%
  \BibitemOpen
  \bibfield  {author} {\bibinfo {author} {\bibfnamefont {B.~E.}\ \bibnamefont {Breckner}}, \bibinfo {author} {\bibfnamefont {H.}~\bibnamefont {Lisei}},\ and\ \bibinfo {author} {\bibfnamefont {G.}~\bibnamefont {{Ionuţ Şimon}}},\ }\bibfield  {title} {\bibinfo {title} {Optimal control results for a class of stochastic schrödinger equations},\ }\href {https://doi.org/https://doi.org/10.1016/j.amc.2021.126310} {\bibfield  {journal} {\bibinfo  {journal} {Applied Mathematics and Computation}\ }\textbf {\bibinfo {volume} {407}},\ \bibinfo {pages} {126310} (\bibinfo {year} {2021})}\BibitemShut {NoStop}%
\bibitem [{\citenamefont {Ahn}\ \emph {et~al.}(2002)\citenamefont {Ahn}, \citenamefont {Doherty},\ and\ \citenamefont {Landahl}}]{Ahn_2002}%
  \BibitemOpen
  \bibfield  {author} {\bibinfo {author} {\bibfnamefont {C.}~\bibnamefont {Ahn}}, \bibinfo {author} {\bibfnamefont {A.~C.}\ \bibnamefont {Doherty}},\ and\ \bibinfo {author} {\bibfnamefont {A.~J.}\ \bibnamefont {Landahl}},\ }\bibfield  {title} {\bibinfo {title} {Continuous quantum error correction via quantum feedback control},\ }\bibfield  {journal} {\bibinfo  {journal} {Physical Review A}\ }\textbf {\bibinfo {volume} {65}},\ \href {https://doi.org/10.1103/physreva.65.042301} {10.1103/physreva.65.042301} (\bibinfo {year} {2002})\BibitemShut {NoStop}%
\bibitem [{\citenamefont {Villanueva}\ and\ \citenamefont {Kappen}(2025)}]{bkappen}%
  \BibitemOpen
  \bibfield  {author} {\bibinfo {author} {\bibfnamefont {A.}~\bibnamefont {Villanueva}}\ and\ \bibinfo {author} {\bibfnamefont {H.}~\bibnamefont {Kappen}},\ }\href {https://arxiv.org/abs/2410.18635} {\bibinfo {title} {Stochastic optimal control of open quantum systems}} (\bibinfo {year} {2025}),\ \Eprint {https://arxiv.org/abs/2410.18635} {arXiv:2410.18635 [quant-ph]} \BibitemShut {NoStop}%
\bibitem [{\citenamefont {Street}\ and\ \citenamefont {Crisan}(2021)}]{semimartingale}%
  \BibitemOpen
  \bibfield  {author} {\bibinfo {author} {\bibfnamefont {O.~D.}\ \bibnamefont {Street}}\ and\ \bibinfo {author} {\bibfnamefont {D.}~\bibnamefont {Crisan}},\ }\bibfield  {title} {\bibinfo {title} {Semi-martingale driven variational principles},\ }\href {https://doi.org/10.1098/rspa.2020.0957} {\bibfield  {journal} {\bibinfo  {journal} {Proceedings of the Royal Society A: Mathematical, Physical and Engineering Sciences}\ }\textbf {\bibinfo {volume} {477}},\ \bibinfo {pages} {20200957} (\bibinfo {year} {2021})}\BibitemShut {NoStop}%
\bibitem [{\citenamefont {Welch}(1967)}]{psdwelch}%
  \BibitemOpen
  \bibfield  {author} {\bibinfo {author} {\bibfnamefont {P.}~\bibnamefont {Welch}},\ }\bibfield  {title} {\bibinfo {title} {The use of fast fourier transform for the estimation of power spectra: a method based on time averaging over short, modified periodograms},\ }\href@noop {} {\bibfield  {journal} {\bibinfo  {journal} {IEEE Transactions on audio and electroacoustics}\ }\textbf {\bibinfo {volume} {15}},\ \bibinfo {pages} {70} (\bibinfo {year} {1967})}\BibitemShut {NoStop}%
\bibitem [{\citenamefont {de~Keijzer}\ \emph {et~al.}(2023)\citenamefont {de~Keijzer}, \citenamefont {Tse},\ and\ \citenamefont {Kokkelmans}}]{vqoc2}%
  \BibitemOpen
  \bibfield  {author} {\bibinfo {author} {\bibfnamefont {R.}~\bibnamefont {de~Keijzer}}, \bibinfo {author} {\bibfnamefont {O.}~\bibnamefont {Tse}},\ and\ \bibinfo {author} {\bibfnamefont {S.}~\bibnamefont {Kokkelmans}},\ }\bibfield  {title} {\bibinfo {title} {Pulse based {V}ariational {Q}uantum {O}ptimal {C}ontrol for hybrid quantum computing},\ }\href {https://doi.org/10.22331/q-2023-01-26-908} {\bibfield  {journal} {\bibinfo  {journal} {{Quantum}}\ }\textbf {\bibinfo {volume} {7}},\ \bibinfo {pages} {908} (\bibinfo {year} {2023})}\BibitemShut {NoStop}%
\bibitem [{\citenamefont {Khaneja}\ \emph {et~al.}(2005)\citenamefont {Khaneja}, \citenamefont {Reiss}, \citenamefont {Kehlet}, \citenamefont {Schulte-Herbrüggen},\ and\ \citenamefont {Glaser}}]{GRAPE}%
  \BibitemOpen
  \bibfield  {author} {\bibinfo {author} {\bibfnamefont {N.}~\bibnamefont {Khaneja}}, \bibinfo {author} {\bibfnamefont {T.}~\bibnamefont {Reiss}}, \bibinfo {author} {\bibfnamefont {C.}~\bibnamefont {Kehlet}}, \bibinfo {author} {\bibfnamefont {T.}~\bibnamefont {Schulte-Herbrüggen}},\ and\ \bibinfo {author} {\bibfnamefont {S.~J.}\ \bibnamefont {Glaser}},\ }\bibfield  {title} {\bibinfo {title} {Optimal control of coupled spin dynamics: design of nmr pulse sequences by gradient ascent algorithms},\ }\href {https://doi.org/https://doi.org/10.1016/j.jmr.2004.11.004} {\bibfield  {journal} {\bibinfo  {journal} {Journal of Magnetic Resonance}\ }\textbf {\bibinfo {volume} {172}},\ \bibinfo {pages} {296} (\bibinfo {year} {2005})}\BibitemShut {NoStop}%
\bibitem [{Note1()}]{Note1}%
  \BibitemOpen
  \bibinfo {note} {For readability purposes standard bra-ket notation is replaced by daggers to indicate conjugates, as in Refs.~\cite {semina,ssepaper}}\BibitemShut {NoStop}%
\bibitem [{Note2()}]{Note2}%
  \BibitemOpen
  \bibinfo {note} {For notational purposes we will write $\phi _t$ for $\phi ^\protect \mathbf {z}_t$, where the dependence on $\protect \mathbf {z}$ is implicit}\BibitemShut {NoStop}%
\bibitem [{stn(2002)}]{stnratio}%
  \BibitemOpen
  \bibinfo {title} {Power spectral density analysis},\ in\ \href {https://doi.org/https://doi.org/10.1002/0471439207.ch4} {\emph {\bibinfo {booktitle} {Principles of Random Signal Analysis and Low Noise Design}}}\ (\bibinfo  {publisher} {John Wiley \& Sons, Ltd},\ \bibinfo {year} {2002})\ Chap.~\bibinfo {chapter} {4}, pp.\ \bibinfo {pages} {92--137}\BibitemShut {NoStop}%
\bibitem [{\citenamefont {Young}\ and\ \citenamefont {Whaley}(2012)}]{dephasingnoise}%
  \BibitemOpen
  \bibfield  {author} {\bibinfo {author} {\bibfnamefont {K.~C.}\ \bibnamefont {Young}}\ and\ \bibinfo {author} {\bibfnamefont {K.~B.}\ \bibnamefont {Whaley}},\ }\bibfield  {title} {\bibinfo {title} {Qubits as spectrometers of dephasing noise},\ }\href@noop {} {\bibfield  {journal} {\bibinfo  {journal} {Physical Review A—Atomic, Molecular, and Optical Physics}\ }\textbf {\bibinfo {volume} {86}},\ \bibinfo {pages} {012314} (\bibinfo {year} {2012})}\BibitemShut {NoStop}%
\bibitem [{\citenamefont {Bergli}\ \emph {et~al.}(2009)\citenamefont {Bergli}, \citenamefont {Galperin},\ and\ \citenamefont {Altshuler}}]{frequencynoise}%
  \BibitemOpen
  \bibfield  {author} {\bibinfo {author} {\bibfnamefont {J.}~\bibnamefont {Bergli}}, \bibinfo {author} {\bibfnamefont {Y.~M.}\ \bibnamefont {Galperin}},\ and\ \bibinfo {author} {\bibfnamefont {B.}~\bibnamefont {Altshuler}},\ }\bibfield  {title} {\bibinfo {title} {Decoherence in qubits due to low-frequency noise},\ }\href@noop {} {\bibfield  {journal} {\bibinfo  {journal} {New Journal of Physics}\ }\textbf {\bibinfo {volume} {11}},\ \bibinfo {pages} {025002} (\bibinfo {year} {2009})}\BibitemShut {NoStop}%
\bibitem [{\citenamefont {Maller}\ \emph {et~al.}(2009)\citenamefont {Maller}, \citenamefont {M{\"u}ller},\ and\ \citenamefont {Szimayer}}]{ornstein}%
  \BibitemOpen
  \bibfield  {author} {\bibinfo {author} {\bibfnamefont {R.~A.}\ \bibnamefont {Maller}}, \bibinfo {author} {\bibfnamefont {G.}~\bibnamefont {M{\"u}ller}},\ and\ \bibinfo {author} {\bibfnamefont {A.}~\bibnamefont {Szimayer}},\ }\bibfield  {title} {\bibinfo {title} {Ornstein--uhlenbeck processes and extensions},\ }\href@noop {} {\bibfield  {journal} {\bibinfo  {journal} {Handbook of financial time series}\ ,\ \bibinfo {pages} {421}} (\bibinfo {year} {2009})}\BibitemShut {NoStop}%
\bibitem [{\citenamefont {Bensoussan}(2018)}]{Bensoussan_2018}%
  \BibitemOpen
  \bibfield  {author} {\bibinfo {author} {\bibfnamefont {A.}~\bibnamefont {Bensoussan}},\ }\href@noop {} {\emph {\bibinfo {title} {Estimation and control of Dynamical Systems}}}\ (\bibinfo  {publisher} {Springer},\ \bibinfo {year} {2018})\BibitemShut {NoStop}%
\bibitem [{\citenamefont {Peng}(1993)}]{backwardsde1}%
  \BibitemOpen
  \bibfield  {author} {\bibinfo {author} {\bibfnamefont {S.}~\bibnamefont {Peng}},\ }\bibfield  {title} {\bibinfo {title} {Backward stochastic differential equations and applications to optimal control},\ }\href {https://doi.org/10.1007/BF01195978} {\bibfield  {journal} {\bibinfo  {journal} {Applied Mathematics and Optimization}\ }\textbf {\bibinfo {volume} {27}},\ \bibinfo {pages} {125} (\bibinfo {year} {1993})}\BibitemShut {NoStop}%
\bibitem [{\citenamefont {Jr.}\ \emph {et~al.}(1996)\citenamefont {Jr.}, \citenamefont {Ma},\ and\ \citenamefont {Protter}}]{backwardsde2}%
  \BibitemOpen
  \bibfield  {author} {\bibinfo {author} {\bibfnamefont {J.~D.}\ \bibnamefont {Jr.}}, \bibinfo {author} {\bibfnamefont {J.}~\bibnamefont {Ma}},\ and\ \bibinfo {author} {\bibfnamefont {P.}~\bibnamefont {Protter}},\ }\bibfield  {title} {\bibinfo {title} {{Numerical methods for forward-backward stochastic differential equations}},\ }\href {https://doi.org/10.1214/aoap/1034968235} {\bibfield  {journal} {\bibinfo  {journal} {The Annals of Applied Probability}\ }\textbf {\bibinfo {volume} {6}},\ \bibinfo {pages} {940 } (\bibinfo {year} {1996})}\BibitemShut {NoStop}%
\bibitem [{\citenamefont {van Dijk}\ \emph {et~al.}(2019)\citenamefont {van Dijk}, \citenamefont {Kawakami}, \citenamefont {Schouten}, \citenamefont {Veldhorst}, \citenamefont {Vandersypen}, \citenamefont {Babaie}, \citenamefont {Charbon},\ and\ \citenamefont {Sebastiano}}]{electricfields}%
  \BibitemOpen
  \bibfield  {author} {\bibinfo {author} {\bibfnamefont {J.}~\bibnamefont {van Dijk}}, \bibinfo {author} {\bibfnamefont {E.}~\bibnamefont {Kawakami}}, \bibinfo {author} {\bibfnamefont {R.}~\bibnamefont {Schouten}}, \bibinfo {author} {\bibfnamefont {M.}~\bibnamefont {Veldhorst}}, \bibinfo {author} {\bibfnamefont {L.}~\bibnamefont {Vandersypen}}, \bibinfo {author} {\bibfnamefont {M.}~\bibnamefont {Babaie}}, \bibinfo {author} {\bibfnamefont {E.}~\bibnamefont {Charbon}},\ and\ \bibinfo {author} {\bibfnamefont {F.}~\bibnamefont {Sebastiano}},\ }\bibfield  {title} {\bibinfo {title} {Impact of classical control electronics on qubit fidelity},\ }\href {https://doi.org/10.1103/PhysRevApplied.12.044054} {\bibfield  {journal} {\bibinfo  {journal} {Phys. Rev. Appl.}\ }\textbf {\bibinfo {volume} {12}},\ \bibinfo {pages} {044054} (\bibinfo {year} {2019})}\BibitemShut {NoStop}%
\bibitem [{Note3()}]{Note3}%
  \BibitemOpen
  \bibinfo {note} {Note that this does not imply $X_{l,t}=W_{l,t}$, instead $X_{l,t}$'s only random source is $W_{l,t}$, e.g. Ornstein-Uhlenbeck noise \cite {Barchielli_2010}}\BibitemShut {NoStop}%
\bibitem [{\citenamefont {Daele}(1997)}]{haarmeasure}%
  \BibitemOpen
  \bibfield  {author} {\bibinfo {author} {\bibfnamefont {A.}~\bibnamefont {Daele}},\ }\bibfield  {title} {\bibinfo {title} {The haar measure on finite quantum groups},\ }\href@noop {} {\bibfield  {journal} {\bibinfo  {journal} {Proceedings of the American Mathematical Society}\ }\textbf {\bibinfo {volume} {125}},\ \bibinfo {pages} {3489} (\bibinfo {year} {1997})}\BibitemShut {NoStop}%
\bibitem [{\citenamefont {de~L\'es\'eleuc}\ \emph {et~al.}(2018)\citenamefont {de~L\'es\'eleuc}, \citenamefont {Barredo}, \citenamefont {Lienhard}, \citenamefont {Browaeys},\ and\ \citenamefont {Lahaye}}]{parameters1}%
  \BibitemOpen
  \bibfield  {author} {\bibinfo {author} {\bibfnamefont {S.}~\bibnamefont {de~L\'es\'eleuc}}, \bibinfo {author} {\bibfnamefont {D.}~\bibnamefont {Barredo}}, \bibinfo {author} {\bibfnamefont {V.}~\bibnamefont {Lienhard}}, \bibinfo {author} {\bibfnamefont {A.}~\bibnamefont {Browaeys}},\ and\ \bibinfo {author} {\bibfnamefont {T.}~\bibnamefont {Lahaye}},\ }\bibfield  {title} {\bibinfo {title} {Analysis of imperfections in the coherent optical excitation of single atoms to rydberg states},\ }\href {https://doi.org/10.1103/PhysRevA.97.053803} {\bibfield  {journal} {\bibinfo  {journal} {Phys. Rev. A}\ }\textbf {\bibinfo {volume} {97}},\ \bibinfo {pages} {053803} (\bibinfo {year} {2018})}\BibitemShut {NoStop}%
\bibitem [{\citenamefont {Wudarski}\ \emph {et~al.}(2023)\citenamefont {Wudarski}, \citenamefont {Zhang}, \citenamefont {Korotkov}, \citenamefont {Petukhov},\ and\ \citenamefont {Dykman}}]{parameters3}%
  \BibitemOpen
  \bibfield  {author} {\bibinfo {author} {\bibfnamefont {F.}~\bibnamefont {Wudarski}}, \bibinfo {author} {\bibfnamefont {Y.}~\bibnamefont {Zhang}}, \bibinfo {author} {\bibfnamefont {A.~N.}\ \bibnamefont {Korotkov}}, \bibinfo {author} {\bibfnamefont {A.}~\bibnamefont {Petukhov}},\ and\ \bibinfo {author} {\bibfnamefont {M.}~\bibnamefont {Dykman}},\ }\bibfield  {title} {\bibinfo {title} {Characterizing low-frequency qubit noise},\ }\href {https://doi.org/10.1103/PhysRevApplied.19.064066} {\bibfield  {journal} {\bibinfo  {journal} {Phys. Rev. Appl.}\ }\textbf {\bibinfo {volume} {19}},\ \bibinfo {pages} {064066} (\bibinfo {year} {2023})}\BibitemShut {NoStop}%
\bibitem [{\citenamefont {Jones}\ \emph {et~al.}(2019)\citenamefont {Jones}, \citenamefont {Endo}, \citenamefont {McArdle}, \citenamefont {Yuan},\ and\ \citenamefont {Benjamin}}]{spectrum}%
  \BibitemOpen
  \bibfield  {author} {\bibinfo {author} {\bibfnamefont {T.}~\bibnamefont {Jones}}, \bibinfo {author} {\bibfnamefont {S.}~\bibnamefont {Endo}}, \bibinfo {author} {\bibfnamefont {S.}~\bibnamefont {McArdle}}, \bibinfo {author} {\bibfnamefont {X.}~\bibnamefont {Yuan}},\ and\ \bibinfo {author} {\bibfnamefont {S.~C.}\ \bibnamefont {Benjamin}},\ }\bibfield  {title} {\bibinfo {title} {Variational quantum algorithms for discovering hamiltonian spectra},\ }\href {https://doi.org/10.1103/PhysRevA.99.062304} {\bibfield  {journal} {\bibinfo  {journal} {Phys. Rev. A}\ }\textbf {\bibinfo {volume} {99}},\ \bibinfo {pages} {062304} (\bibinfo {year} {2019})}\BibitemShut {NoStop}%
\bibitem [{\citenamefont {Polonyi}\ and\ \citenamefont {Rachid}(2021)}]{liouvillian}%
  \BibitemOpen
  \bibfield  {author} {\bibinfo {author} {\bibfnamefont {J.}~\bibnamefont {Polonyi}}\ and\ \bibinfo {author} {\bibfnamefont {I.}~\bibnamefont {Rachid}},\ }\bibfield  {title} {\bibinfo {title} {Elementary open quantum states},\ }\bibfield  {journal} {\bibinfo  {journal} {Symmetry}\ }\textbf {\bibinfo {volume} {13}},\ \href {https://doi.org/10.3390/sym13091624} {10.3390/sym13091624} (\bibinfo {year} {2021})\BibitemShut {NoStop}%
\bibitem [{\citenamefont {Dellacherie}\ and\ \citenamefont {Meyer}(1978)}]{filtration}%
  \BibitemOpen
  \bibfield  {author} {\bibinfo {author} {\bibfnamefont {C.}~\bibnamefont {Dellacherie}}\ and\ \bibinfo {author} {\bibfnamefont {P.}~\bibnamefont {Meyer}},\ }\href {https://books.google.nl/books?id=K75ZNAEACAAJ} {\emph {\bibinfo {title} {Probabilities and Potential}}},\ \bibinfo {series} {North-Holland mathematics studies}\ No.\ \bibinfo {number} {v. 1}\ (\bibinfo  {publisher} {Hermann},\ \bibinfo {year} {1978})\BibitemShut {NoStop}%
\bibitem [{\citenamefont {Song}\ and\ \citenamefont {Wang}(2021)}]{Song2021}%
  \BibitemOpen
  \bibfield  {author} {\bibinfo {author} {\bibfnamefont {J.}~\bibnamefont {Song}}\ and\ \bibinfo {author} {\bibfnamefont {M.}~\bibnamefont {Wang}},\ }\bibfield  {title} {\bibinfo {title} {Stochastic maximum principle for systems driven by local martingales with spatial parameters},\ }\href {https://api.semanticscholar.org/CorpusID:235293783} {\bibfield  {journal} {\bibinfo  {journal} {Probability, Uncertainty and Quantitative Risk}\ } (\bibinfo {year} {2021})}\BibitemShut {NoStop}%
\bibitem [{\citenamefont {Kunita}\ and\ \citenamefont {Watanabe}(1967)}]{martingalerepresentation}%
  \BibitemOpen
  \bibfield  {author} {\bibinfo {author} {\bibfnamefont {H.}~\bibnamefont {Kunita}}\ and\ \bibinfo {author} {\bibfnamefont {S.}~\bibnamefont {Watanabe}},\ }\bibfield  {title} {\bibinfo {title} {On square integrable martingales},\ }\href {https://api.semanticscholar.org/CorpusID:118173938} {\bibfield  {journal} {\bibinfo  {journal} {Nagoya Mathematical Journal}\ }\textbf {\bibinfo {volume} {30}},\ \bibinfo {pages} {209 } (\bibinfo {year} {1967})}\BibitemShut {NoStop}%
\bibitem [{\citenamefont {Bayram}\ \emph {et~al.}(2018)\citenamefont {Bayram}, \citenamefont {Partal},\ and\ \citenamefont {Orucova~Buyukoz}}]{stochasticintegration1}%
  \BibitemOpen
  \bibfield  {author} {\bibinfo {author} {\bibfnamefont {M.}~\bibnamefont {Bayram}}, \bibinfo {author} {\bibfnamefont {T.}~\bibnamefont {Partal}},\ and\ \bibinfo {author} {\bibfnamefont {G.}~\bibnamefont {Orucova~Buyukoz}},\ }\bibfield  {title} {\bibinfo {title} {Numerical methods for simulation of stochastic differential equations},\ }\href {https://doi.org/10.1186/s13662-018-1466-5} {\bibfield  {journal} {\bibinfo  {journal} {Advances in Difference Equations}\ }\textbf {\bibinfo {volume} {2018}},\ \bibinfo {pages} {17} (\bibinfo {year} {2018})}\BibitemShut {NoStop}%
\bibitem [{\citenamefont {Kloeden}\ and\ \citenamefont {Platen}(1999)}]{platen}%
  \BibitemOpen
  \bibfield  {author} {\bibinfo {author} {\bibfnamefont {P.~E.}\ \bibnamefont {Kloeden}}\ and\ \bibinfo {author} {\bibfnamefont {E.}~\bibnamefont {Platen}},\ }\href@noop {} {\emph {\bibinfo {title} {Numerical solution of stochastic differential equations}}}\ (\bibinfo  {publisher} {Springer},\ \bibinfo {year} {1999})\BibitemShut {NoStop}%
\end{thebibliography}%

\onecolumngrid

\appendix

\section{Matrix transformations}
\label{app:matrices}
Our quantity of interest is the fidelity $\F:=|\phi^\dagger \psi|^2$, as the overlap between $\phi$ (the desired state without noise) and the noisy state $\psi$ evolving according to \eqref{eq:sse}. To derive an explicit formula for $\F$ a system of real-valued SDEs for a vector $\bta\in \mathbb{C}^m$, $m\ge 1$, is derived where $\bta_t=[{\phi_t}^\dagger P_0 \psi_t, {\phi_t}^\dagger P_1 \psi_t,...,{\phi_t}^\dagger P_{4^N-1}\psi_t]$. Here $N$, is the number of qubits and $P_i$ is one of the $4^N$ individual Pauli matrices. Fixing $P_0=I^{\otimes N}$ ensures ${\bta_t}^\dagger \Lambda_0 \bta_t = F_t$, with $\Lambda_0=[1,0,0,...,0][1,0,0,...,0]^\dagger$. The system of equations for $\bta$ is given by
\begin{equation}
\label{eq:ssesystem2}
\begin{aligned}
    d\bta_t&=\sum_j z_j(t) A_j \bta_t dt-\frac{1}{2}\sum_l \gamma_l^2 B^\dagger_l B_l \bta_t dt +\sum_l \gamma_l B_l \bta_t dX_{l,t}\\
    &=: b(t,\bta,z)dt+ \sum_l \sigma_l(t,\bta,z) dX_{l,t},
\end{aligned}
\end{equation}
where the anti-Hermitian matrices $A_j$ and $B_j$ have elements 
\begin{equation}
\begin{aligned}
    (A_j)_{m,n}&=i\Tr[P_m[H_j,P_n]]=-\overline{(A_j)_{n,m}},\\
    (B_l)_{m,n}&=i\Tr[P_mP_nS_l]=-\overline{(B_l)_{n,m}}.
\end{aligned}    
\end{equation}
From \eqref{eq:ssesystem}, one can show that ${\bta_t}^\dagger \bta_t={\bta_0}^\dagger \bta_0=2$ is a conserved quantity. As $x_{0,0}=1$, the rest of the initial state can be seen as a point on the $4^N-1$ dimensional sphere. In certain cases, it is easier to consider the variable $Q:=\bta\bta^\dagger$ which evolves according to \begin{equation}
    dQ=\sum_j z_j(t) (A_j Q-QA_j^\dagger)dt+\frac{1}{2}\sum_l\gamma_l^2(-2B_l Q B_l^\dagger +B_l^\dagger B_l Q +Q B_l^\dagger B_l)dt +\sum_l \gamma_l(B_l Q-QB_l^\dagger)dX_{l,t}
\end{equation}
which through the Fock-Liouville isomorphism \cite{liouvillian} is equivalent to 
\begin{equation}
\label{eq:odesystem}   
\di\V=\mathbf{A}\V\, \di t+\sum_l\mathbf{B}_l\V\, \di X_{l,t}+\mathbf{a}\,\di t+\sum_l\mathbf{b}_l\,\di X_{l,t},
\end{equation}
where constant entries of $\V$ have been incorporated in the inhomogeneous part $\mathbf{a}, \mathbf{b}$. Note that the first component of $\V$ now directly corresponds to the fidelity $\F=|\phi^\dagger\psi|^2$. The size of the system varies depending on the properties of the noise operator $S$.

\section{Derivation of Gâteaux derivative}
\label{app:ssederivation}

Here, we establish the derivation of the Gâteaux derivative $\nabla_z J_3$ term in~\eqref{eq:costfunction}. We want to find the gradient of 
\begin{equation}
\label{eq:costfunctionapp}
J_3(\z)= -\mu\mathbb{E}\left[{\bta_t}^\dagger\Lambda_0 \bta_t+\nu\int_0^T {\bta_s}^\dagger \Lambda_0 \bta_s d s \right]
\end{equation}
where $x$ evolves under the system of controlled stochastic differential equations given by
\begin{equation}
\label{eq:ssesystemapp}
    d\bta_t=\sum_j z_j(t) A_j \bta_t \di t-\frac{1}{2}\sum_l \gamma_{l}^2|z_{c(l)}(t)|^q B^\dagger_l B_l \bta_t d[X]_{l,t} +\sum_l \gamma_l|z_{c(l)}(t)|^{q/2} B_l \bta_t dX_{l,t}. 
\end{equation}
Setting $q=0$ or $q=1$, respectively, results in the fixed noise and scaled noise versions. We denote the filtration of all combined noise processes $X_{l,t}$ as $\mathcal{F}_t$ \cite{filtration}. For this derivation, we combine the approaches in Ref.~\cite[Ch.\ 11]{Bensoussan_2018} and Ref.~\cite{Song2021}, which generalizes the results in Ref.~\cite{Bensoussan_2018} from white noise processes to general colored noise processes.

According to Ref.~\cite{Song2021}, the Gâteaux derivative can be expressed as 
\begin{equation}
\begin{aligned}
\label{eq:derivativeapp}    \nabla J(\z)&[\delta z_j]=\mathbb{E}\left[ \int_0^T K_{z_j}(\bta_t, z(t), p(t), r(t)) \delta z_j(t) d t\right]\\
    &=\mathbb{E} \left[\int_0^T \left(p^\dagger(t) A_j \bta_t-\sum_{l|c(l)=j} \left(\frac{q}{2} \gamma_l^2 p^\dagger(t) B^\dagger_l B_l \bta_t+ \frac{q}{2} \frac{\gamma_l}{|z_j(t)|^{q/2}}  x^\dagger(t) B^\dagger_l r_l(t)\right)\right)\delta z_j(t) \di t\right]\\
    &= \mathbb{E}\left[ \int_0^T  p^\dagger(t) A_j \bta_t \delta z_j(t) \di t\right]+\sum_{l|c(l)=j}    \mathbb{E}\left[\int_{0}^T - \frac{q}{2} \gamma_l^2 p^\dagger(t) B^\dagger_l B_l \bta_t\delta z_j(t)\di t \right]+\mathbb{E}\left[\int_0^T\frac{q}{2} \frac{\gamma_l}{|z_j(t)|^{q/2}}  r^\dagger_l(t) B_l \bta_t\delta z_j(t) \di t \right]\\
    &=: \nabla J_{3,a}(\z)[\delta z_j]+\nabla J_{3,b}(\z)[\delta z_j]+\nabla J_{3,c}(\z)[\delta z_j] 
\end{aligned}
\end{equation}
where $r_l(t):=\gamma_l |z_{c(l)}(t)|^{q/2} \tilde{r}(t) B_l \bta_t$, and  $p$ and $\tilde{r}$ are the unique adjoint processes satisfying 
\begin{equation}
\begin{aligned}
-d p(t)&= \left(\sum_j z_j(t) {A_j}^\dagger p+\sum_{l}\gamma_l^2|z_{c(l)}(t)|^q B^\dagger_l \tilde{r}(t)B_l \bta_t+\Lambda_0\bta_t\right) d t-\tilde{r}(t)\sum_l\gamma_l|z_{c(l)}(t)|^{q/2} B_l \bta_t d X_{l,t}+d N_t\\
&=\left(\sum_j z_j(t) {A_j}^\dagger p+\sum_{l}\gamma_l|z_{c(l)}(t)|^{q/2} B^\dagger_l r_l(t)+\Lambda_0\bta_t\right) d t-\sum_l r_l(t) d X_{l,t}+d N_t, \quad p(T)= \Lambda_0\bta_t.
\end{aligned}
\end{equation}
According to Ref.~\cite{Song2021}, the solution triplet $(p,\tilde{r},N)$  is uniquely determined. Here $N$ is a mean-zero local martingale orthogonal to all white noise processes generating the noises $X_{l,t}$. As we assume our noise processes are fully generated by white noise processes, this can only be true if $N=0$. 

From Ref.~\cite{Bensoussan_2018}, we find
\begin{equation}
\begin{aligned}
    p(t) &= -\nu\Psi^\dagger(t) \int_0^t \Phi^\dagger(s) \Lambda_0 \bta_s d s+\Psi^\dagger(t) \mathbb{E}\left[\zeta(0)| \mathcal{F}^t\right],\\
r_l(t) &= \Psi^\dagger(t) G_l(t)-\gamma_l |z_l(t)|^{q/2} B^\dagger_l p(t),\\
\zeta(t) &= \Phi^\dagger(T) \Lambda_0\bta_t + \nu\int_t^T \Phi^\dagger(s) \Lambda_0 \bta_s d s, 
\end{aligned}
\end{equation}

where $\Phi$ and $\Psi$ are solutions to the forward stochastic differential equations given by
\begin{equation}
\begin{aligned}
d \Phi(t) &= \sum_j z_j(t) A_j \Phi(t) \di t-\frac{1}{2}\sum_l \gamma_l^2 |z_{c(l)}(t)|^q B^\dagger_l B_l \Phi(t) \di t +\sum_{l} \gamma_l |z_{c(l)}(t)|^{q/2}B_l\Phi(t) dX_{l,t},\quad\Phi(0) =I,\\
-d \Psi(t)& =\sum_j z_j(t)\Psi(t) A_j \di t+\frac{1}{2}\sum_l \gamma_l^2 |z_{c(l)}(t)|^q \Psi(t) B^\dagger_l B_l \di t+\Psi(t) \sum_{l} \gamma_l
|z_{c(l)}(t)|^{q/2}B_l d X_{l,t}, \quad\Psi(0) =I,
\end{aligned}
\end{equation}
and $G_l$ are martingales defined using the representation theorem of martingales \cite{martingalerepresentation} by
\begin{equation}
\mathbb{E}\left[\zeta^\dagger(0) \mid \mathcal{F}^t\right]-\mathbb{E}\left[\zeta^\dagger(0) \mid \mathcal{F}^0\right]=\sum_{l} \int_0^t G_l(s) dW_{l,s}.
\end{equation}
Filling this into~\eqref{eq:derivativeapp}  yields
\begin{equation}
\begin{aligned}
    \nabla_{z} J_{3,a}[\delta z_j]&=    \mathbb{E}\left[ \int_0^T  p^\dagger(t) A_j \bta_t \delta z_j(t) \di t\right]\\
    &=\mathbb{E}\left[\int_0^T  \left(-\nu\Psi^\dagger(t) \int_0^t \Phi^\dagger(s) \Lambda_0 x(s) d s+\Psi^\dagger(t) \mathbb{E}\left[\zeta^\dagger(0) \mid \mathcal{F}^t\right]\right)^\dagger A_j \bta_t \delta z_j(t)\right]\\
    &=\mathbb{E}\left[\int_0^T \zeta(t) \Psi(t) A_j \bta_t \delta z_j(t)\di t\right]\\
\end{aligned}    
\end{equation}
and
\begin{equation}
\begin{aligned}
    \nabla_{z} J_{3,b}[\delta z_j]+\nabla_{z} J_{3,c}[\delta z_j]=& \sum_{l|c(l)=j}    \mathbb{E}\left[\int_{0}^T - \frac{q}{2} \gamma_l^2 p^\dagger(t) B^\dagger_l B_l \bta_t\delta z_j(t)\di t \right]+\mathbb{E}\left[\int_0^T\frac{q}{2} \frac{\gamma_l}{|z_j(t)|^{q/2}}  r^\dagger_l(t) B_l \bta_t\delta z_j(t) \di t \right]\\
    =&\sum_{l|c(l)=j}  \mathbb{E}\left[\int_{0}^T - \frac{q}{2} \gamma_l^2 p^\dagger(t) B^\dagger_l B_l \bta_t\delta z_j(t)\di t \right]+\mathbb{E}\left[\int_{0}^T  \frac{q}{2} \gamma_l^2 p^\dagger(t) B^\dagger_l B_l \bta_t\delta z_j(t)\di t \right]\\
    &+\mathbb{E}\left[\int_0^T\frac{q}{2} \frac{\gamma_l}{|z_j(t)|^{q/2}} {G_l}^\dagger(t) \Psi(t) B_l \bta_t\delta z_j(t) \di t \right]\\
    &=\mathbb{E}\left[\sum_{l'}  \int_0^T {G_{l'}}^\dagger(t) dW_{{l'},t} \sum_{l|c(l)=j}  \int_0^T \frac{q}{2} \frac{\gamma_l}{|z_j(t)|^{q/2}} \Psi(t) B_l \bta_t\delta z_j(t) dW_{l,t} \right]\\
    &=\mathbb{E}\left[\big(\mathbb{E}[\zeta(0)|\mathcal{F}_T]-\mathbb{E}[\zeta(0)|\mathcal{F}_0]\big) \sum_{l|c(l)=j}  \int_0^T \frac{q}{2} \frac{\gamma_l}{|z_j(t)|^{q/2}} \Psi(t) B_l \bta_t\delta z_j(t) dW_{l,t} \right]\\
    &=\mathbb{E}\left[\int_0^T \sum_{l|c(l)=j}  \zeta(0) \Psi(t) \frac{q}{2} \frac{\gamma_l}{|z_j(t)|^{q/2}} B_l \bta_t\delta z_j(t) dW_{l,t} \right]\\
    &=\mathbb{E}\left[\int_0^T \sum_{l|c(l)=j}  \zeta(t) \Psi(t) \frac{q}{2} \frac{\gamma_l}{|z_j(t)|^{q/2}} B_l \bta_t\delta z_j(t) dW_{l,t} \right]
\end{aligned}    
\end{equation}
Combining these terms gives
\begin{equation}
\begin{aligned}
    \nabla_{z} J_{3}[\delta z_j]&= \nabla_{z} J_{3,a}[\delta z_j]+ \nabla_{z} J_{3,b}[\delta z_j]+ \nabla_{z}J_{3,c}[\delta z_j]\\
    &=\mathbb{E}\left[\int_{0}^T \zeta(t)\Psi(t)\left(A_j \di t+\sum_{l|c(l)=j}\frac{q}{2}\frac{\gamma_l}{|z_j(t)|^{q/2}}B_ldW_{l,t}\right)\bta_t\delta z_j(t)\right]
\end{aligned}
\end{equation}

\section{Stochastic Integration}
\label{app:stochint}
Numerical calculation of the gradients is performed using stochastic integration. Throughout this work, stochastic differential equations of the form
\begin{equation}
    \di \Y=a(\Y)\,\di t+b(\Y)\,\di W_t,
\end{equation}
are solved using the explicit (weak) second-order scheme due to Platen \cite{stochasticintegration1,platen}. This scheme is given by 
\begin{equation}
\begin{aligned}
& \Y_{n+1}=\Y_n+\frac{1}{2}\big(a(\bar{\Upsilon})+a(\Y_n)\big) \Delta t+\frac{1}{4}\big(b\left(\bar{\Upsilon}^{+}\right)+b\left(\bar{\Upsilon}^{-}\right)+2 b(\Y_n)\big) \mathcal{N}\sqrt{\Delta t} +\frac{1}{4}\big(b(\tilde{\Upsilon}^{+})-b(\widetilde{\Upsilon}^{-})\big)\left(\mathcal{N}^2-1\right) \sqrt{\Delta t}, 
\end{aligned}
\end{equation}
with supporting values $\bar{\Upsilon}=\Y_n+a(\Y_n) \Delta t+b(\Y_n) \mathcal{N}\sqrt{\Delta t}$ and $\bar{\Upsilon}^{ \pm}=\Y_n+a(\Y_n) \Delta t \pm b(\Y_n) \sqrt{\Delta t}$. Here $\mathcal{N}$ is a sample from a standard normal distribution and $\Delta t>0$ is a time step. Heuristically, for our types of problems, this scheme leads to better convergence than standard Euler-Maruyama methods \cite{stochasticintegration1}, most likely due to non-Euclidity of the Hilbert space and non-Lipschitz behavior of colored noise \cite{ssepaper}.

\end{document}